\documentclass[a4paper,12pt]{article}
\pdfoutput=1
\usepackage{graphicx,rotating,hyperref,slashed,amsmath,xcolor,amssymb,amsfonts,expdlist,colortbl,cite,expdlist,youngtab}
\makeatletter
\hypersetup{colorlinks,bookmarksopen,bookmarksnumbered,
linkcolor=blus,pdfstartview=FitH,urlcolor=rossos,citecolor=verde}
\allowdisplaybreaks

\def\lsim{\mathrel{\rlap{\lower3pt\hbox{\hskip0pt$\sim$}}
   \raise1pt\hbox{$<$}}}         
\def\gsim{\mathrel{\rlap{\lower4pt\hbox{\hskip1pt$\sim$}}
   \raise1pt\hbox{$>$}}}         

 \newcommand{\ecm}{e\,{\rm cm}}
 \newcommand{\sfootnote}[1]{}
\definecolor{bluc}{cmyk}{1,1,0,0.1}
\definecolor{rossoCP3}{cmyk}{0,.88,.77,.40}
\definecolor{rosso}{cmyk}{0,1,1,0.4}
\definecolor{rossos}{cmyk}{0,1,1,0.55}
\definecolor{rossoc}{cmyk}{0,1,1,0.2}
\definecolor{verdes}{cmyk}{0.92,0,0.59,0.4}

\hypersetup{colorlinks, bookmarksopen, bookmarksnumbered,
citecolor=verdes, linkcolor=bluc, pdfstartview=FitH, urlcolor=rossos}

 \renewcommand{\Im}{{\rm Im}\,}

\newcommand{\mio}[1]{}

\newcommand{\xxx}[1]{{\color{red}\bf[#1]}}

\definecolor{Gray}{gray}{0.95}

\newcommand{\F}{{\cal F}}
\newcommand{\Q}{{\cal F}}
\renewcommand{\S}{{\cal S}}

\newcommand{\sfrac}[2]{#1/#2}

\newcommand{\LTC}{\Lambda_{\rm TC}}

\usepackage{multicol}
\usepackage{color}
\definecolor{rosso}{cmyk}{0,1,1,0.4}
\definecolor{rossos}{cmyk}{0,1,1,0.55}
\definecolor{rossoc}{cmyk}{0,1,1,0.2}
\definecolor{blu}{cmyk}{1,1,0,0.3}
\definecolor{blus}{cmyk}{1,1,0,0.6}
\definecolor{bluc}{cmyk}{1,1,0,0.1}
\definecolor{verde}{cmyk}{0.92,0,0.59,0.25}
\definecolor{verdec}{cmyk}{0.92,0,0.59,0.15}
\definecolor{verdes}{cmyk}{0.92,0,0.59,0.4}

\newcommand{\eq}[1]{~{\rm (\ref{eq:#1})}}

\newcommand{\GeV}{\,{\rm GeV}}
\newcommand{\TeV}{\,{\rm TeV}}

\newcommand{\Tr}{\,{\rm Tr}}
\newcommand{\diag}{\,{\rm diag}}

\def\circa#1{\,\raise.3ex\hbox{$#1$\kern-.75em\lower1ex\hbox{$\sim$}}\,}

\newcommand{\beq}{\begin{equation}}
\newcommand{\eeq}{\end{equation}}

\newcommand{\bea}{\begin{eqnarray}}
\newcommand{\eea}{\end{eqnarray}}
\newcommand{\be}{\begin{equation}}
\newcommand{\ee}{\end{equation}}
\font\tenrsfs=rsfs10 at 12pt
\font\sevenrsfs=rsfs7 at 10 pt
\font\fiversfs=rsfs5
\newfam\rsfsfam
\textfont\rsfsfam=\tenrsfs
\scriptfont\rsfsfam=\sevenrsfs
\scriptscriptfont\rsfsfam=\fiversfs
\def\mathscr#1{{\fam\rsfsfam\relax#1}}

\def\Lag{\mathscr{L}}

\def\circa#1{\,\raise.3ex\hbox{$#1$\kern-.75em\lower1ex\hbox{$\sim$}}\,}
\makeatletter

\def\hhref#1{\href{http://arxiv.org/abs/#1}{arXiv:#1}} 

\newcommand{\doi}[1]{\href{http://dx.doi.org/#1}{[doi]}}

\def\hhref#1{\href{http://arxiv.org/abs/#1}{arXiv:#1}} 
 
\def\art{\@ifnextchar[{\eart}{\oart}}
\def\eart[#1]#2#3#4#5#6{{\rm #2}, {\em #3 \bf #4} {\rm (#6) #5} ({\em #1})}

\def\article{\@ifnextchar[{\earticle}{\oarticle}}
\def\oarticle#1#2#3#4#5#6{{\rm #1}, {\em ``#6''}, {\rm #2 #3 (#5) #4}}
\def\earticle[#1]#2#3#4#5#6#7{{\rm #2}, {\em ``#7''}, {\rm #3 #4 (#6) #5}  [\hhref{#1}]}
\def\hepart[#1]#2{{\rm #2, \em#1}}
\def\heparticle[#1]#2#3{#2, {\em ``#3''} [\hhref{#1}]}

\newcounter{alphaequation}[equation]
\def\thealphaequation{\theequation\hbox to
0.6em{\hfil\alph{alphaequation}\hfil}}
\def\eqnsystem#1{
\def\@eqnnum{{\rm (\thealphaequation)}}
\def\@@eqncr{\let\@tempa\relax \ifcase\@eqcnt \def\@tempa{& & &} \or
  \def\@tempa{& &}\or \def\@tempa{&}\fi\@tempa
  \if@eqnsw\@eqnnum\refstepcounter{alphaequation}\fi
\global\@eqnswtrue\global\@eqcnt=0\cr}
\refstepcounter{equation} \let\@currentlabel\theequation \def\@tempb{#1}
\ifx\@tempb\empty\else\label{#1}\fi
\refstepcounter{alphaequation}
\let\@currentlabel\thealphaequation
\global\@eqnswtrue\global\@eqcnt=0 \tabskip\@centering\let\\=\@eqncr
$$\halign to \displaywidth\bgroup \@eqnsel\hskip\@centering
$\displaystyle\tabskip\z@{##}$&\global\@eqcnt\@ne
\hskip2\arraycolsep\hfil${##}$\hfil& \global\@eqcnt\tw@\hskip2\arraycolsep
$\displaystyle\tabskip\z@{##}$\hfil
\tabskip\@centering&\llap{##}\tabskip\z@\cr}
\def\endeqnsystem{\@@eqncr\egroup$$\global\@ignoretrue} \makeatother

\newcommand{\SU}{\,{\rm SU}}
\newcommand{\SO}{\,{\rm SO}}
\newcommand{\Sp}{\,{\rm Sp}}
\newcommand{\U}{\,{\rm U}}

\definecolor{fiorentina}{rgb}{.5,0,.5}

\begin{document}
\centerline{CERN-PH-TH-2016-148\hfill  CP3-Origins-2016-027 \hfill EFI-16-15 \hfill IFUP-TH/2016}

\vspace{1truecm}

\begin{center}
\boldmath

{\textbf{\LARGE\color{rossoCP3} Fundamental partial compositeness}}
\unboldmath

\bigskip\bigskip

\vspace{0.1truecm}

{\bf Francesco Sannino$^a$,
Alessandro Strumia$^{b,c}$, Andrea Tesi$^{d}$, Elena Vigiani$^b$}
 \\[8mm]
{\it $^a$ CP$^3$-Origins and Danish IAS, University of Southern Denmark, Campusvej 55, Denmark}\\[1mm]
{\it $^b$ Dipartimento di Fisica dell'Universit{\`a} di Pisa and INFN, Italy}\\[1mm]
{\it $^c$ CERN, Theory Division, Geneva, Switzerland}\\[1mm]
{\it $^d$ Department of Physics, Enrico Fermi Institute, University of Chicago, Chicago, IL 60637}\\[2mm]

\vspace{1cm}

\thispagestyle{empty}
{\large\bf\color{blus} Abstract}
\begin{quote}\large
We construct renormalizable Standard Model extensions, valid up to the Planck scale, 
that give a composite Higgs from a new fundamental strong force acting on fermions  and scalars.
Yukawa interactions of these particles with Standard Model fermions realize the partial compositeness scenario.
Under certain assumptions on the dynamics of the scalars, successful models exist because gauge quantum numbers of Standard Model fermions
admit a minimal enough `square root'.
Furthermore, right-handed SM fermions have an $\SU(2)_R$-like structure, yielding a custodially-protected composite Higgs.
Baryon and lepton numbers arise accidentally. Standard Model fermions acquire mass at tree level, 
while the Higgs potential and flavor violations are generated by quantum corrections.
We further discuss accidental symmetries and other dynamical features stemming from the new strongly interacting scalars. 
If the same phenomenology can be obtained from models without our elementary scalars,
 they would reappear as composite states.
\end{quote}
\thispagestyle{empty}
\end{center}

\setcounter{page}{1}
\setcounter{footnote}{0}

\newpage

\tableofcontents

\section{Introduction}
Is the Higgs boson elementary or composite?  It is often argued that elementary scalars cannot be light in the absence
of a mechanism that protects their masses from quantum corrections. A time-honoured solution is to make scalars emerge from new composite dynamics featuring fermions.   A 
pseudo-Nambu-Goldstone boson Higgs with a compositeness scale below a TeV is therefore considered natural.
However a consistent TeV-scale composite dynamics able to reproduce the successes of the Standard Model (SM) Higgs in the flavor sector resulted in an unresolved challenge. 

This situation prompted theorists to focus on effective field theories that supposedly capture the low energy manifestation of some unknown underlying strongly-coupled dynamics, especially in the flavor sector.
As it is well known from pion physics, accidental symmetries of the underlying strong dynamics provide important insight on the low energy effective theory. Composite Higgs effective Lagrangians postulate ad hoc symmetries and features that allow to be consistent with data, that are compatible with the predictions of an elementary Higgs. However {\it cosettology} (assumptions about global symmetries) and resulting effective field theories do not guarantee the existence of an underlying fundamental composite dynamics.  
 
Nevertheless a relevant composite paradigm is currently being under intensive study and it is based on three main hypotheses. First, the Higgs is part of a weak doublet of  pseudo-Goldstone bosons ~\cite{Kaplan:1983fs,Kaplan:1983sm,Dugan:1984hq}. The original idea has been investigated via effective descriptions more recently in~\cite{Agashe:2004rs,Giudice:2007fh}.\footnote{ De facto, effective descriptions cannot discern a composite realization from a more economical elementary Goldstone Higgs realization \cite{Alanne:2014kea,Gertov:2015xma}.} Second, extra custodial symmetries are added to ameliorate compatibility with experimental bounds (see \cite{Panico:2015jxa} for a review). Third, fermion masses are reproduced by postulating {\it partial compositeness}  \cite{Kaplan:1991dc} --- namely that each SM fermion $f$ acquires mass by mixing with an heavier composite fermion (see also \cite{pc-5d}). 

According to the partial compositeness prescription, each SM fermion $f$ couples linearly to a composite fermionic operator ${\cal B}$  through an interaction of the form $f\,{\cal B}$. Large anomalous dimensions\footnote{Anomalous dimensions are physical quantities only in presence of (near) conformal dynamics.} of the operator ${\cal B}$ (typically composed by several fermionic fields)  are then invoked such that the operator $ f  {\cal B}$ is either  super-renormalizable or marginal. However recent studies of the anomalous dimensions of conformal baryon operators in $\SU(3)$ gauge theories suggest that it is hard to achieve the required very large anomalous dimensions in purely fermionic theories~\cite{Pica:2016rmv}. 

One might consider highly involved models or hope that behind these attempts there might exist yet unknown strongly coupled dynamics
possibly stemming from warped extra-dimensional scenarios or from
exotic CFTs that do not have a four dimensional quantum field theoretical description,
giving rise to scalar operators ${\cal O}_\S$ with dimension $\dim{\cal O}_\S \simeq 1$ (in order to reproduce data),
and with $\dim {\cal O}_\S^2 $ larger than 4 (in order to avoid naturalness issues)~\cite{hep-ph/0409274}.
General considerations exclude this possibility~\cite{0807.0004,1109.5176}.   
The simplest option is then that ${\cal O}_\S$ is just an elementary scalar $\S$.
Any theory that mimics an elementary scalar is presumably more simply described by an explicit  elementary scalar.

Because of the challenges above, we investigate here extensions of the SM featuring a composite Higgs sector made by a new fundamental techni-strong  theory that besides featuring techni-fermions ($\cal F$) also features techni-scalars ($\cal S$)\footnote{Our construction  differs from {\it bosonic technicolor} \cite{bosonicTC,Antola:2009wq}  where a TC-singlet elementary Higgs is added to the composite TC-fermion dynamics. 
A TC-colored scalar was introduced  for the top quark in eq.~(2.6) of~\cite{hep-ph/9807469}, 
but in the context of TC dynamics that breaks the SM, while we consider a composite Higgs and fermion partial compositeness.
One can even try to naturalize these theories by supersymmetrizing them~\cite{Antola:2010nt}, or simply take them at the face value of alternative models of electroweak symmetry breaking~\cite{StrongNat}. }. We introduce the techni-scalars primarily to construct composite techni-baryons $ \cal B =  F  S$  and associate linear interactions with the SM fermions in order to successfully implement the partial compositeness paradigm. In fact, by construction the new composite techni-baryons have mass dimensions close to the minimum required of $5/2$.  Because any purely fermionic extension~\cite{gherghetta, ferrettiPC, vecchi2, ferretti} is required to have composite baryons with dimensions close to $5/2$, these baryons would presumably behave as if they were made by a fermion and a composite scalar similar to ours (see also~\cite{serone} for a supersymmetric realization). 

\begin{figure}
$$\includegraphics[width=0.95\textwidth]{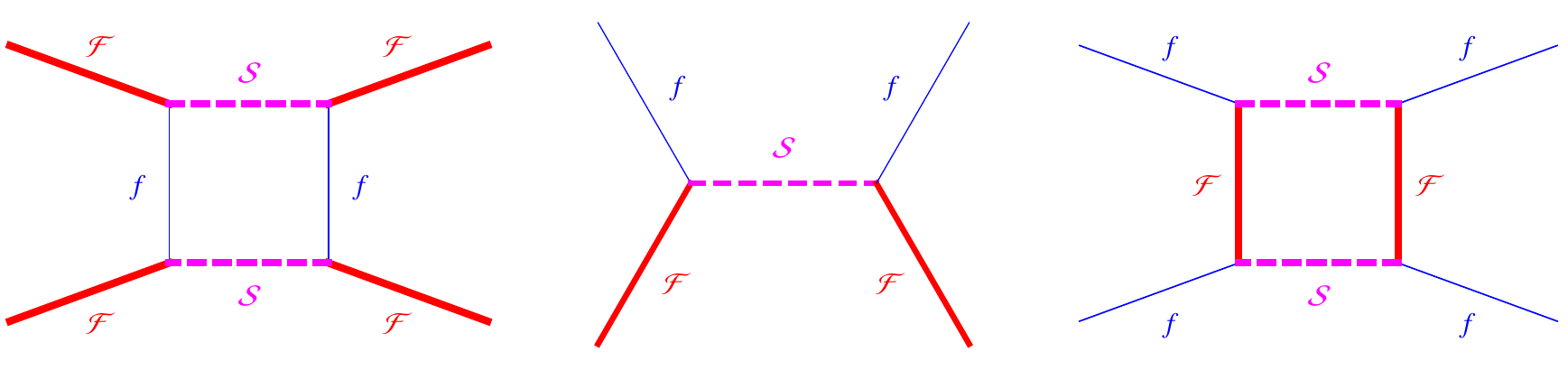}$$
\vspace{-1cm}
\caption{\em \label{flavour} We here denote as $f$  the SM fermions, as $\F$ the TC-fermions and as $\S$ the TC-scalars.
The diagram on the left contributes to $\F^4$ operators, which become the Higgs potential, given that $H\sim \F\F$
(extra TC-penguin diagrams are not shown).
The tree-level diagram in the middle leads to $\F^2 f^2$ operators, which become the SM $ffH$ Yukawa couplings.
The diagram on the right contributes to $f^4$ operators, which give corrections to flavor physics.
Each diagram gets dressed by strong TC interactions.
}
\end{figure}

The technicolor group is indicated by  $G_{\rm TC}$, which can be either $\SU(N)$, $\SO(N)$ or $\Sp(N)$  with vectorial techni-fermions and techni-scalars. For historical reasons we will use the technicolor (TC) terminology for the underlying composite dynamics.  

We will choose the TC-particle content such that it  automatically leads to a custodial symmetry, as well as accidental conservation of baryon and lepton number, like in the SM.
Partial compositeness is realized provided that the gauge quantum numbers allow
each fundamental SM fermion 
to have a fundamental Yukawa coupling to at least one pair of TC particles:
\beq\label{eq:Yuk}
\hbox{(each SM fermion)} \times \hbox{(some TC scalar)} \times \hbox{(some TC fermion).}\eeq
Figure~\ref{flavour} illustrates how these Yukawa couplings lead to SM fermions masses at tree level, 
as well as to an Higgs potential and to new flavor violations at loop level.
TC-fermions and TC-scalars acquire specific patterns of accidental global symmetries, spontaneously broken by the TC dynamics:
the Higgs can be identified with a light techni-pion (TC$\pi$) made either of two TC-fermions or of two TC-scalars.
We further constrain the SM {extensions} to avoid  sub-Planckian Landau poles and require the TC model to lead to chiral symmetry breaking.

The key to the success is to find a TC gauge group and associated TC-fermions and TC-scalars with  appropriate SM quantum numbers. In practice this requires satisfying eq.\eq{Yuk} by finding a minimal `square root' of SM fermions gauge quantum numbers.  We will show that it is possible to construct successful composite Higgs theories and associated partially composite sectors. 
Despite the presence of TC-scalars,
the Higgs mass is calculable if $H$ is a pseudo-Goldstone boson made of two TC-fermions.

%

\medskip

In section~\ref{models} we discuss general issues about the strong dynamics of scalars, 
which has not been studied outside the special case of supersymmetry.
We present the separate pieces that must be combined together in succesfull concrete models,  
which are presented in section~\ref{models2} (the eager phenomenologist might want to jump here). 
In section~\ref{SMYuk} we compute the resulting Higgs physics.
We present our conclusions in section~\ref{end}.

\section{General preliminary considerations}\label{models}

We consider a  theory with gauge group $G_{\rm TC}\otimes G_{\rm SM}$ 
and vectorial TC-fermions and TC-scalars in the fundamental of $G_{\rm TC}$ with UV Lagrangian 
\beq 
\Lag = \Lag_{\rm SM}^{H=0}  +\Lag_{\rm kin}+\Lag_Y -V \,,
\eeq
where $\Lag_{\rm kin}$ contains kinetic, gauge interactions, and possible masses $m_\F$ and $m_\S$ for the TC particles.

We use a compact notation for the particle spectrum quantum numbers under the SM gauge group $G_{\rm SM}=\SU(3)_c\otimes\SU(2)_L\otimes\U(1)_Y$  that we exemplify via the SM fermion fields:
\begin{equation}
L=(1,2)_{-1/2},~E=(1,1)_1, ~Q=(3,2)_{1/6},~U=(\bar 3,1)_{-2/3},~D=(\bar 3,1)_{1/3},~N=(1,1)_0\ . 
\end{equation} 
We define as $R^c$ the representation conjugated to $R$, e.g.\ $U^c= (3,1)_{2/3}$.

We indicate Weyl  TC-fermions by $\Q$ and generically complex TC-scalars by $\S$. They will decompose under $G_{\rm SM}$ as: 
\beq \Q = \Q_L\oplus  \Q_E\oplus  \Q_Q \oplus  \Q_U\oplus  \Q_D\oplus  \Q_N\oplus\cdots,\qquad
\S= \S_L\oplus  \S_E\oplus  \S_Q \oplus  \S_U\oplus  \S_D\oplus  \S_N\oplus\cdots\eeq 
with, for example,
\beq \Q_L =(1,2)_{-1/2} \ , \qquad  \S_Q=(3,2)_{1/6}  \qquad {\rm under}~G_{\rm SM}.
\eeq
Clearly each $\Q$ and $\S$ field carries a further gauge index  under $G_{\rm TC}$ that we omitted. 

In the following we will consider for $G_{\rm TC}$ either $\SU(N)$, $\SO(N)$ or $\Sp(N)$ gauge groups and we will assume TC-fermions and TC-scalars to live in the fundamental of these groups,
which minimize the contributions to gauge $\beta$ functions, as needed for successful models.
We will consider TC-fermions vector-like with respect to $G_{\rm TC}$ (in the $\SU(N)_{\rm TC}$ case for each TC-fermion $\Q$ in the fundamental there will be also $ \Q^c$ in the anti-fundamental) and to $G_{\rm  SM}$.

\subsection{Accidental global symmetries}\label{sec:global}
We now classify the global symmetries of a given TC theory for different choices of $G_{\rm TC}$, once the SM interactions are switched off. 
\begin{itemize}
\item $G_{\rm TC}=\SU(N)_{\rm TC}$ for $N>2$ has a complex fundamental.
The vectorial TC-fermions $\Q$ can be organized in
terms of Dirac spinors $\Psi_{\Q} = (\Q,\bar \Q^c)^T$ and 
the kinetic term can be written as
\beq\label{eq:Lkin}
\Lag_{\rm kin}=- \frac14 {\cal G}_{\mu\nu}^2 + \overline \Psi_{\Q} (i \slashed{D}- m_{\Q})\Psi_{\Q}+
( | D_\mu \S|^2 - m_ \S ^2 | \S |^2)  \,,
\eeq
where the sum over color and flavor indices is understood.
Ignoring the SM gauge interactions,
the fermionic kinetic term has a `TC flavor' non-anomalous global symmetry 
$\SU(N_F)_L \otimes \SU(N_F)_R \otimes \U(1)_V$, 
where $N_F$ is the dimension of the SM representation to which $\Q$ belongs to, that  for $\Q = \Q_E \oplus \Q_L + \cdots$ is $N_F = \dim L + \dim E + \cdots$. 
 
In the scalar sector, the kinetic term of TC scalars
similarly has a $\SU(N_S) \otimes \U(1)_S$ global symmetry where $N_S$ counts the number of complex scalars in the fundamental of TC, that for $\S = \S_E \oplus \S_L + \cdots$ is $N_S= \dim L + \dim E + \cdots$.

\item 
$G_{\rm TC}=\SO(N)_{\rm TC}$ has a vectorial real representation\footnote{Spinorial matter representations
of $\SO(N)$  contribute less to $\beta$ functions than fundamental representations only for $N\le 6$;
these are already taken into account since $\SO(6)\sim \SU(4)$,
$\SO(5)\sim\Sp(4)$, $\SO(4)\sim\SU(2)^2$, $\SO(3)\sim\SU(2)\sim\Sp(2)$.
We explore fundamentals of $\SO(N)$ which do not correspond to fundamentals of the equivalent groups.} and therefore 
Weyl spinors  $\F$ in the fundamental of TC must lie in a real representation of $G_{\rm  SM}$.
The global symmetry  is
 $\SU(N_F)$ with $N_F$ the dimension of the real SM representation to which $\Q$ belongs to.
 
In the scalar sector, we define $N_S$ as the number of real copies of $N$:
for example $N_S=6$ for a TC-scalar in the fundamental 3 of $\SU(3)_c$ and $N_S=3$ for a TC-scalar in the 3 of $\SU(2)_L$.
The scalar kinetic term 
$(D_\mu  \S _i)^T (D_\mu  \S ^i)$ 
has accidental global symmetry 
$\SO(N_S) \otimes \mathbb Z_2$. 

\medskip

\item 
$G_{\rm TC}=\Sp(N)_{\rm TC}$ with even $N$ is defined as the group of matrices that leave invariant the antisymmetric tensor $\gamma =\varepsilon \otimes \mathbb I_{N/2}$
where $\varepsilon= i \sigma_2$ is the 2-dimensional antisymmetric tensor.
The fundamental of $\Sp(N)$ is pseudo-real (see appendix~\ref{Sp}).
Again, we consider vectorial TC-fermions  $\Q $ constructing vectorial SM representations, with $N_F$ Weyl fermions in the fundamental of $\Sp(N)_{\rm TC}$. 
$N_F$ counts the dimension of the real SM representation of $\Q$ and it must be even to avoid the Witten topological anomaly.
As for the orthogonal TC gauge group the fermion kinetic term 
has the non-abelian global symmetry $\SU(N_F)$. 

In the scalar sector, the kinetic term of $N_S$ complex scalars in the $N$ of $\Sp(N)_{\rm TC}$ has accidental global symmetry $\Sp(2N_S)$, see appendix~\ref{Sp}. 
For example, a scalar in the $(3,N)$ of $\SU(3)_c\otimes \Sp(N)_{\rm TC}$ has $N_S=3$ and global symmetry $\Sp(6)$.


\end{itemize}
The global symmetries of the kinetic terms are summarized in table \ref{tab:symmetries}.\footnote{Supersymmetric models
predict extra Yukawa and quartic couplings of order $g_{\rm TC}$, such a unique global symmetry acts on fermions and scalars.}

\begin{table}\small
$$\begin{array}{|c|c|ccccc|}\hline
\hbox{Fields}& \hbox{Gauge} & \multicolumn{3}{c|}{\hbox{Global symmetry of fermions}} &  \multicolumn{2}{c|}{\hbox{Global, scalars}}\\
\hline  \hline  
\rowcolor[cmyk]{0,0.1,0,0}
     & \SU(N)_{\rm TC} & \SU(N_F)_L& \SU(N_F)_R  & \U(1)_V & \SU(N_S) & \U(1)_S \\ 
\hline  
\rowcolor[cmyk]{0,0.1,0,0} {\mathcal F} & N & N_F& 1 & +1 & 1 & 0  \\
\rowcolor[cmyk]{0,0.1,0,0} {\mathcal F}^c & \bar N & 1 &\bar N_F & -1 & 1 & 0  \\\rowcolor[cmyk]{0,0.1,0,0} {\mathcal S} &  N & 1 &1 & 0 & N_S & 1  \\
\hline \hline
\rowcolor[cmyk]{0.1,0,0,0}     & \SO(N)_{\rm TC} && \SU(N_F) 
      & & \mathrm{O}(N_S) & \\
\hline 
\rowcolor[cmyk]{0.1,0,0,0} {\mathcal F} & N && N_F & &  1 & \\
\rowcolor[cmyk]{0.1,0,0,0} {\mathcal S} & N && 1  & &  N_S & \\
\hline \hline
\rowcolor[cmyk]{0,0,0.25,0}     & \Sp(N)_{\rm TC} && \SU(N_F) 
     &  &   \Sp(2 N_S)  & \\
\hline 
\rowcolor[cmyk]{0,0,0.25,0}{\mathcal F} & N && N_F &  &   1 & \\
\rowcolor[cmyk]{0,0,0.25,0}{\mathcal S} & N && 1  & &   2 N_S &
\\ \hline
\end{array}$$
\caption{\em\label{tab:symmetries}  Gauge and local non anomalous symmetries of generic models
with gauge group $G_{\rm TC}=\SU(N)_{\rm TC}$ or $\SO(N)_{\rm TC}$ or $\Sp(N)_{\rm TC}$ and
 $N_F$ ($N_S$) TC-fermions (TC-scalars) in the $N$ representation;
for $\SU(N)_{\rm TC}$ and $\Sp(N)_{\rm TC}$ we count complex scalars, for $\SO(N)_{\rm TC}$ we count real scalars. 
}
\end{table}

\bigskip
 
Group theory allows to construct renormalizable Yukawa operators of the form of eq.\eq{Yuk}.
(For $N=3$
Yukawa interactions among 3 TC particles are also possible).
These are the operators leading naturally to the partial compositeness scenario. In fact when the techni-force is strong enough it will create the fermionic bound state ${\cal B= F S}$ that already has mass dimension 5/2 at the engineering level. Also, we do not need an extra mechanism or additional force to construct the  overall ${f \cal B}$ operator. Furthermore,  since any other construction for partial compositeness will have to yield a composite fermion with at most mass dimension 5/2 we expect that at the  effective description level it will reduce to our construction. 
The simplest example is a TC-baryon emerging from an $\SU(3)_{\rm TC}$  gauge theory with fundamental TC-fermions. In this case, at the fundamental level, the TC-baryon will be made by three TC-fermions that can always be represented as a bound state of one TC-fermion and  a TC-scalar with the quantum numbers of di-techniquarks. It is a simple matter to show that this intermediate dynamical description can be generalized  to the case in which a TC-baryon is made by TC-fermions in multiple TC representations.\footnote{If TC-scalars result from composite fermionic dynamics the intermediate description  must abide a number of consistent conditions, known as compositeness conditions, that have been properly re-discussed and extended in \cite{Krog:2015bca} for a general class of gauge-Yukawa theories} Of course, in the purely fermionic case, one must argue for the existence of near conformal non-supersymmetric quantum field theories yielding baryon operators with unplausibily large \cite{Pica:2016rmv} anomalous dimensions.

\subsection{Quartic couplings among TC-scalars}\label{TCS?}
TC-scalars develop self-interactions  generated by RGE effects via, for example, their gauge interactions. These effects are encoded in $\beta$ functions  $\beta_\lambda=d\lambda/d\ln E$  which at one loop assume the generic form $(4\pi)^2\beta_\lambda \sim +\lambda^2+g^4_{\rm TC}  - \lambda g^2_{\rm TC} $. Here  we indicated the generic scalar self-couplings by $\lambda$ and the TC gauge coupling with $g_{\rm TC}$. 
Running down to low energy, when the TC gauge coupling start becoming strong, quartics become of order $\pm g^2_{\rm TC}$, where the sign depends on the specific model.
This means that, if quartics remain positive up to the confinement scale, they also contribute to the nonperturbative dynamics of the theory. In this case, a simplifying assumption is to use flavor universal quartics in order not to spoil the symmetries of the scalar sector listed in Table \ref{tab:symmetries}. If, on the other hand, quartics become negative at some energy scale, the Coleman-Weinberg mechanism can take place.\footnote{Although this is not the main focus of this work it is worth mentioning that, 
if a Coleman-Weinberg phenomenon occurs, one can obtain an elementary Goldstone Higgs, as discussed in appendix~\ref{altro}.}

\begin{figure}[t]
$$\includegraphics[width=0.43\textwidth]{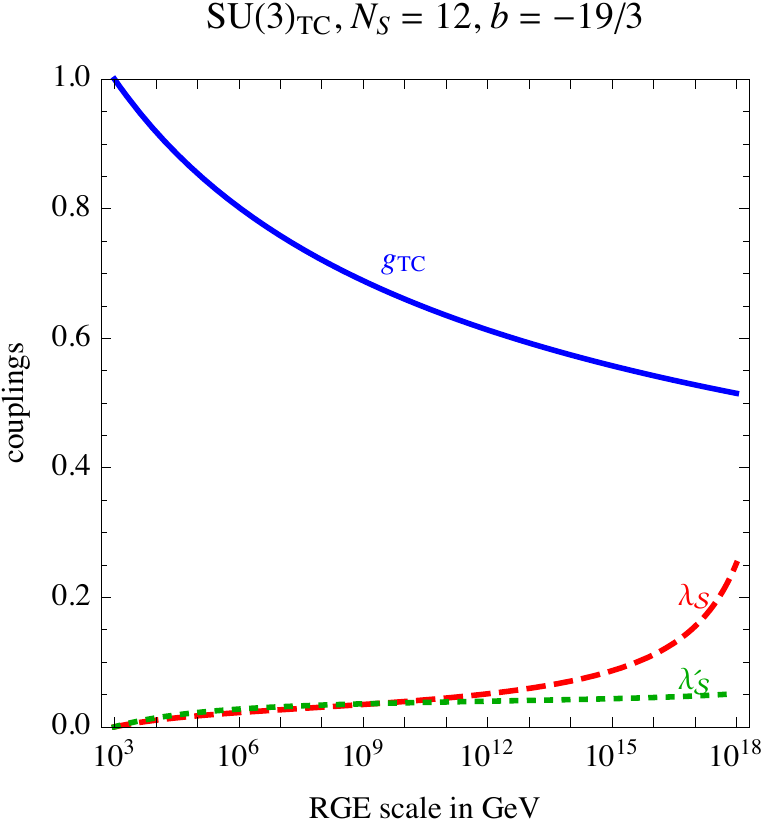}\qquad
\includegraphics[width=0.46\textwidth]{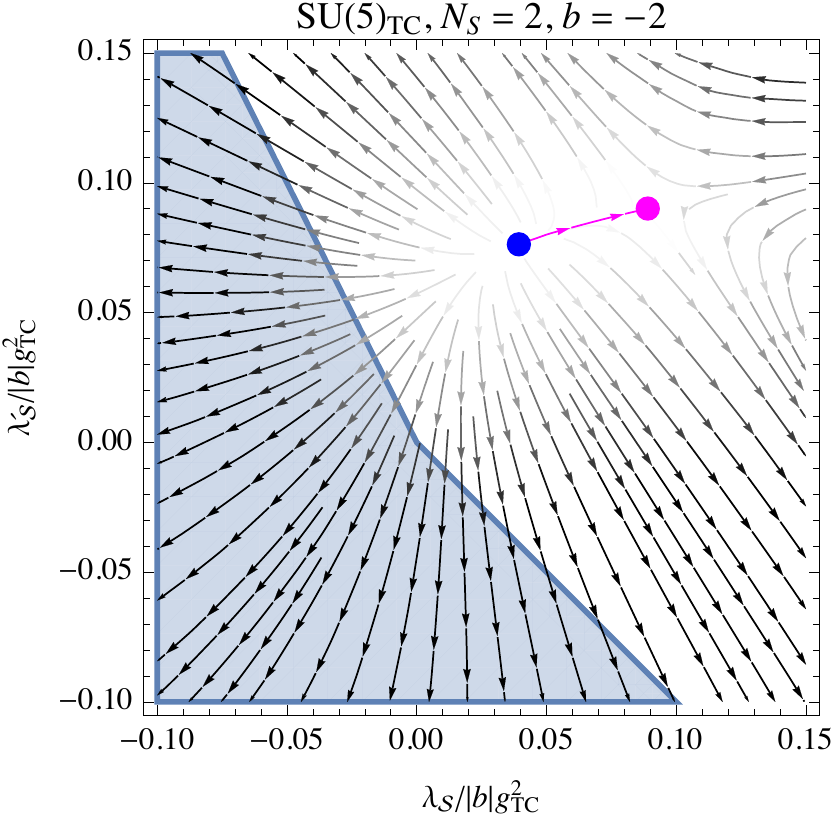}$$ 
\caption{\em \label{fig:Run2} Left: numerical solution to the RGE in the  model of section~\ref{5frags2}.
Right: pseudo-fixed point structure of a model with 2 pseudo-fixed points (shown as dots);
the scalar potential becomes unstable when quartics flow into the shaded region.
}
\end{figure}

To estimate the effects on the quartic self-couplings we consider $N_S$ scalars in the fundamental of $\SU(N)_{\rm TC}$,
such that TC-scalars $\S$ form a complex $N\times N_S$ matrix.
At very high energies, because the couplings are assumed to be small, we can ignore masses and cubic interactions, 
and we write the following quartic potential  including only the $\SU(N_S)$ flavor-symmetric operators
\beq \label{eq:VS}
V=\lambda_\S\Tr(\S\S^\dagger)^2+\lambda'_\S \Tr(\S \S ^\dagger \S \S ^\dagger). \eeq
Such potential is definite positive for $\lambda_{\S } + r \lambda'_{\S } > 0$ where
$r = \Tr(\S \S ^\dagger \S \S ^\dagger) /\Tr(\S \S ^\dagger)^2$ ranges between $1/\min(N,N_S)$ and $1$. 
The relevant one-loop RGEs are
\begin{eqnsystem}{sys:RGE}
(4\pi)^2 \beta_{g_{\rm TC}}&=&bg_{\rm TC}^3\\
(4\pi)^2 \beta_{\lambda_\S } &=&   4 (NN_S+4) \lambda_\S ^2 +12\lambda_\S '^2+ \\
&&
+\lambda_\S \bigg[ 8(N+N_S)\lambda_\S ' - \frac{6(N^2-1)}{N} g^2_{\rm TC} \bigg]+ \frac{3 (N^2 + 2)}{4N^2} g^4_{\rm TC} ,\nonumber \\
(4\pi)^2 \beta_{\lambda_\S '} &=&
4(N+N_S)\lambda_\S '^2+\lambda_\S ' \bigg[24\lambda_\S  - \frac{6(N^2-1)}{N} g^2_{\rm TC}\bigg]+\frac{3 (N^2 - 4)}{4 N}g^4_{\rm TC}.
\end{eqnsystem}
The left panel of fig.~\ref{fig:Run2} shows a sample numerical solution in the model that
will be proposed in section~\ref{5frags2}: quartic couplings can remain numerically small up to the Planck scale.
The right panel shows the pseudo-fixed point structure of a different model that 
admits two pseudo-fixed points (namely, $\lambda/g^2_{\rm TC}$ flow to constant values)
with positive values of the quartics, which can flow
from one point to the other.
 The pseudo-fixed-point conditions 
 can be solved analytically in general~\cite{TAF}
and acquire a simple form in the large $N$ limit, 
where the  equations for the two quartics basically decouple,
showing that pseudo-fixed points $\lambda\sim g^2_{\rm TC}\sim 1/N$
exist
for $N_S< 2N +b+b^2/12N$.

The overall conclusion is that techni-quartic interactions can be well defined till the Planck scale (either with or without interacting pseudo-fixed points) and furthermore that there are theories in which the low energy physics is driven by the techni-gauge interactions becoming strong.

\begin{table}[t]
$$\begin{array}{c|cc}
\hbox{Gauge group} & \hbox{Fermion bilinear condensate}   &\hbox{Intact scalar symmetries} \\ \hline
\SU(N)_{\rm TC} & \SU(N_F)_L\otimes \SU(N_F)_R\to \SU(N_F) &\U(N_S)  \\
\SO(N)_{\rm TC} & \SU(N_F) \to \SO(N_F) &  \mathrm{O}(N_S) \\
\Sp(N)_{\rm TC} & \SU(N_F)\to \Sp(N_F) & \Sp(2 N_S) 
\end{array}$$
\caption{\em\label{tab:G2H}   Pattern of chiral symmetry breaking induced by fermion condensates  $\F \F^c$  for  $\SU(N)_{\rm TC}$ and  $\Q \Q$ for $\SO(N)_{\rm TC}$ and  $\Sp(N)_{\rm TC}$.  An explicit positive squared mass term for the scalars ensures that the scalar symmetries remain intact.}
\end{table}

\subsection{Dynamical symmetry breaking}
\label{sec:condensates}

The analysis above allows us to assume the new strong interaction to be  asymptotically free and  further require that the force is sufficiently strong to confine the fundamental degrees of freedom into techni-hadrons at a scale  $\Lambda_{\rm TC}\circa{>}$~TeV.
Asymptotic freedom is realized when the first order coefficient of the gauge $\beta$ function is negative.
We fill further impose stronger conditions under which it is reasonable to expect  that the underlying dynamics does not display large distance conformality  \cite{hep-ph/0611341,0902.3494}. We will limit here to investigate condensation phenomena induced by the techni-strong force that leave intact the TC gauge interactions.\footnote{We refrain from using very early results~\cite{FS} that would lead to too naive estimates of certain quantities in theories with TC-scalars.}

\subsubsection*{Fermion condensates}
We start by reviewing the pattern of chiral symmetry breaking expected to occur when fermion-bilinear Lorentz preserving condensates form. 
%

In asymptotically free theories with only vector-like fermions one can show \cite{VafaWitten}   that the associated condensates preserve the gauge group
(e.g.\  only gauge-singlets $\langle \bar {\psi}_L \psi_R \rangle$ can form
in $\SU(N)_{\rm TC}$ theories).
Furthermore, it is often argued that non-vanishing fermion condensates orient in such a way
to preserve as much as possible of the original global symmetry provided the massless spectrum is compatible with  the t'Hooft anomaly conditions and other relevant constraints.  
The pattern of symmetry breaking implied by such arguments is summarized in table~\ref{tab:G2H}.
This means that for $G_{\rm TC}=\SU(N)_{\rm TC}$ the TC$\pi$ are made of $ \Q\Q^c$ states (with $\Q$ a Dirac fermion) and for 
$\SO(N)_{\rm TC}$ and $\Sp(N)_{\rm TC}$ are made of $\Q\Q$ (with $\Q$ a Weyl fermion).

\subsubsection*{Scalar condensates}
In the presence of scalars, various dynamical phases can occur. A detailed dynamical study goes beyond the scope of this work, where we just describe the possible patterns of chiral symmetry breaking in the scalar sector stemming from scalar bilinears. The fact that both scalar and fermion bilinears form is partially supported by the intuition gained via the naive most attractive channels approach (MAC)~\cite{DRS}  that here we take merely as guidance. Furthermore, because of the anti-commutative nature of composite operators $\bar{\Q}  \S$,  they cannot acquire a vacuum expectation value and therefore Lorentz invariance is preserved.

An interesting class of models is the one in which the fermions condense and break their global symmetries while the scalars do not break neither their global symmetries nor the gauge interactions. This can be achieved  by endowing the scalars with a positive mass squared  respecting the gauge and scalar global symmetries. In this case the symmetries of the theory are presented in  table~\ref{tab:G2H}. For the scalars to still be actively participating in the TC dynamics we require the explicit common scalar masses to be of the order or smaller than the TC dynamical scale,
$m_\S\circa{<}\Lambda_{\rm TC}$.

Depending on the mass squared term and the strong dynamics one could also have partial spontaneous symmetry breaking of some of their global symmetries and Higgsing of gauge interactions. One of these possibilities will be summarized in table~\ref{tab:vev}.  Alternatively, scalars could break their global symmetries while still leaving confinement intact. 
Scalar condensates proportional to the unity matrix in flavor space do not explicitly break the global accidental symmetry in the scalar sector. This breaking arises if scalars develop flavor non-universal condensates. For example, TC-scalars with the quantum numbers of $L$ and $N$
and a $\langle \S_N\S^*_N\rangle $ condensate  would give rise to the pattern of global symmetry breaking $\U(3) \to \U(2)$ leading to a Goldstone boson $\S_L \S_N^*$ with the quantum numbers of a Higgs doublet plus a singlet.

\smallskip

\subsubsection*{Naive conformal window}

If the number of matter fields of the theory is sufficiently large, and in absence of a Higgsing phenomenon, the TC dynamics
can develop an infrared interacting fixed point. In this scenario no dynamical scale forms at low energies, before coupling the theory to the SM. Since we want the fermions to condense, we must lie outside the conformal window. We provide a crude estimate of the `safe' region of number of fermions and scalars as a function of the number of TC ($N$) where we expect dynamical condensation to occur. 

The one loop $\beta$ function for $g_{\rm TC}$ is:
\be
\label{betaTC}
\beta_{\rm TC}^{(1)} = \frac{g_{\rm TC}^3}{(4 \pi)^2} \left[ -\frac{11}{3} C_2({\rm Adj})+\frac23 T(F) m(F) + \frac13 T(S) m(S) \right] \,,
\ee
where $F$ and $S$ denote the Weyl fermion and complex scalar representations respectively and $m(F)$, $m(S)$ their multiplicity.\footnote{We adopt the common notation $T^a_R T^a_R = C_2(R) \mathbb I$ and ${\rm Tr}[T^a_R T^b_R] = T(R) \delta^{ab}$ with $T_R$ the generators of the $R$ representation.}
In the $\SU(N)_{\rm TC}$ case $m(F)=2 N_F$ and $m(S)=N_S$, for $\SO(N)_{\rm TC}$ we have $m(F)=N_F$ and $m(S)=N_S/2$, while for $\Sp(N)_{\rm TC}$, $m(F)=N_F$ and $m(S)=N_S$ with $N_F$ and $N_S$ defined in section \ref{sec:global} for each choice of $G_{\rm TC}$.

We simply assume that condensates are formed if the first coefficient of the full beta function $\beta^{(1)}_{\rm TC}$, in modulus, is  larger than the third of the modulus of the first coefficient of the gauge beta function, i.e. $\beta_{\rm TC}^{(1)} \circa{<} \frac13 \beta^{(1)}_{\rm TC} |_{\rm gauge}$.  
This is intuitively reasonable since matter screens the confining gauge interactions and the resulting naive condition is roughly compatible with earlier estimates \cite{hep-ph/0611341,0902.3494,1011.3832,1011.5917,1011.4542}.
Considering TC-fermions and TC-scalars in the fundamental of the gauge group we obtain the following conditions:
\beq\label{eq:N>}
\begin{array}{cccc}
G & C_2({\rm Adj}) & T(F) = T(S) & \hbox{Condensates form if}\\
\SU(N)_{\rm TC} & N & \frac12 & \displaystyle N \circa{>}  \frac{3(4 N_F + N_S)}{44} \\
\SO(N)_{\rm TC} & N -2& 1 & \displaystyle N \circa{>} \frac{3(4 N_F+N_S)}{44} + 2 \\
\Sp(N)_{\rm TC} & \frac12(N+2) & \frac12 & \displaystyle N \circa{>} \frac{3(2 N_F +  N_S)}{22} -2\\
\end{array}\eeq
To clarify the counting of fermions and scalars, we
consider for example a TC-fermion and a TC-scalar in the fundamental of the TC group and in the $2_{-1/2}$ of $\SU(2)_L \otimes \U(1)_Y$, if the TC group is $\SU(N)_{\rm TC}$ we have $N_F = N_S = 2$, if it is $\SO(N)_{\rm TC}$ we have $N_F = N_S = 4$ and if it is $\Sp(N)_{\rm TC}$ we have $N_F = 2 N_S = 4$.

%

\subsection{Custodial symmetry}\label{custodial}
The $T$ parameter agrees with SM predictions and gives a strong bound on the
$|H^\dag D_\mu H|^2$ effective operator which can arise in models where $H$ is composite.
The typical correction is of order 
$\hat T \sim v^2/f_{\rm TC}^2$  such that the experimental bound $|\hat{T}|\circa{<} 2\times10^{-3}$ would imply $f_{\rm TC} \circa{>} 5 \TeV$
and a correspondingly large unnatural correction to the Higgs mass.
This unseen deviation from the SM is much suppressed if the Higgs sector respects a `custodial' symmetry $\SU(2)_L \otimes \SU(2)_R \to \SU(2)_c$~\cite{Custodial}.
In fundamental models such symmetry must be a consequence of the TC-fermion content,
arising as an accidental global symmetry.
Below we list the simplest possibilities that lead to a custodial symmetry.
Considering first the case in which the Higgs is a TC$\pi$ made of two TC-fermions:
\begin{enumerate}

\item The most minimal model is obtained considering
$G_{\rm TC}=\Sp(N)$ with TC-fermions 
$\Q = 2_0 \oplus 1_{1/2} \oplus 1_{-1/2}$ under $\SU(2)_L\otimes \U(1)_Y$.\footnote{These TC-fermions are not fragments of any $\SU(5)$ representation; we will later show that they allow to write all needed Yukawa couplings.}
If their mass differences are much smaller than $\Lambda_{\rm TC}$, 
the TC dynamics respects a global symmetry $\SU(4)$ broken to $\Sp(4)$, leading
to TC$\pi$ in a $(2,2) \oplus (1,1)$ of $\SU(2)_L \otimes \SU(2)_R$, among which we can identify the composite Higgs doublet.
In general, whenever there is one Higgs doublet, 
the custodial symmetry is ensured when the unbroken global symmetry group contains a subgroup $\SO(4) \supset \SU(2)_L \otimes \SU(2)_R$.  Lattice simulations \cite{lattice} find that the pattern of chiral symmetry breaking envisioned above  is indeed achieved
for the minimal $\SU(2) \sim \Sp(2)$ case.

 \item Another minimal possibility arises from
 $\SO(N)_{\rm TC}$ with 
TC-fermions  $\Q = \Q_L\oplus \Q_{L^c}\oplus \Q_N$.
If their mass differences are much smaller than $\Lambda_{\rm TC}$, dynamics
 respects a global symmetry  $ \SU(5)$ spontaneously broken to
$\SO(5)$, delivering TC$\pi$ in the $(1,1)\oplus(2,2)\oplus(3,3)$ of $\SU(2)_L \otimes \SU(2)_R$. 
One must check that the extra scalars, such as the triplet in $\Q_L  \Q_{L^c}$, acquire positive squared masses
and have no vev.

 \item Finally, $\SU(N)_{\rm TC}$ with $N\ge 3$ and TC-fermions 
 $\Q= \Q_L\oplus  \Q_{E^c}\oplus  \Q_N$ 
respects a global symmetry  $\SU(4)_L\otimes \SU(4)_R$ spontaneously broken to
$\SU(4)_V$.  If the mass difference between
 $\Q_{E^c}$, $\Q_N$ 
is much smaller than $\Lambda_{\rm TC}$,
TC-strong dynamics respects a global custodial $\SU(2)_L\otimes\SU(2)_R$ symmetry, under which
the TC$\pi$ in the adjoint of $\SU(4)_V$ transform as
$2\times (2,2) \oplus (1,1) \oplus  (3,1) \oplus (1,3)$.
Unlike in the previous cases, one has a complex bidoublet of Higgses:
in the presence of two Higgs doublets, a generic 
minimum of the potential breaks the electro-weak and the custodial symmetry. 
The vacuum expectation values of the two Higgses must be aligned.
A generic potential can have appropriate minima~\cite{SSVZ}; however
special potentials (such as those arising for TC$\pi$)
can need an extra discrete symmetry in order to obtain the desired alignement~\cite{2HDM}. 
One must check that the extra scalars, such as the triplet in $\Q_L \Q^c_L$, acquire positive squared masses,
and have no vacuum expectation values.

\end{enumerate}
 We next consider the case where the Higgs is a TC$\pi$ made of two TC-scalars, recalling that they can have the
accidental global symmetry listed in table~\ref{tab:G2H}, and that its breaking pattern is model dependent.
We assume that scalar condensates preserve $G_{\rm TC}$ and break the global symmetry as follows:
\begin{itemize}

\item $\SU(N)_{\rm TC}$ with TC-scalars $\S=\S_L\oplus \S_{E^c} \oplus \S_N$ 
respects a global $\SU(4)$ symmetry.
A $G_{\rm SM}$-preserving condensate $\langle \S^*\S\rangle= f^2 {\mathbb I}+f'^{2} \diag(0,0,1,1)$ would break it
into $\SU(2)_L \otimes \SU(2)_R$, leading to two custodially protected Higgs doublets.  The TC$\pi$ decompose as $2 \times (2,2) \oplus (1,1)$ under the $\SU(2)_L \otimes \SU(2)_R$ unbroken symmetry.

\item $\SO(N)_{\rm TC}$ with TC-scalars $\S=\S_L \oplus \S_N$ respects a global $\SO(5)$ symmetry.
The most generic $G_{\rm SM}$-preserving condensate $\langle \S\S\rangle= f^2 {\mathbb I}+ f'^2 \diag(0,0,0,0,1)$ would break it
into $\SO(4)$, leading to one custodially protected Higgs doublet.

\item $\Sp(N)_{\rm TC}$ with TC-scalars $\S=\S_L \oplus \S_N$ respects a global $\Sp(6)$ symmetry.
The 
$G_{\rm SM}$-preserving condensate $\langle \S\S\rangle=\varepsilon\otimes\diag(f^2,f^2,f^2+f^{\prime 2})$  breaks $\Sp(6)\to \Sp(4)\otimes\Sp(2)$, leading to  
8 TC$\pi$ that decompose under $\SU(2)_L \otimes \SU(2)_R$ as $2\times (2,2) $, giving two custodially protected Higgs doublets. 
\end{itemize}

\subsubsection*{Custodial symmetry for $Z\to b\bar b$}
In order to reproduce the large top Yukawa coupling, 
the 3rd-generation $Q=(t_L,b_L)$ must be significantly mixed with composite fermions ${\cal B}$. This can give gauge interactions that deviate from those of an elementary fermion.
Thereby a correction of order $\delta g_{b_L} \sim v^2/\Lambda_{\rm TC}^2 $ 
to the $Zb_L\bar b_L$ coupling would imply $\Lambda_{\rm TC}\circa{>}5\TeV$.
This bound is less severe than the one from the $T$ parameter, but it is serious enough that
various authors have discussed how to alleviate it via effective field theories with extra custodial symmetries ~\cite{continite}.
One such example  needs a left-right symmetry that exchanges $\SU(2)_L$ with $\SU(2)_R$ and that $b_L$ respects the condition $T_L=T_R$ and $T^3_L=T^3_R$ under the custodial group~\cite{continite}. 
This can be realized if the composite spectrum in the top sector is $LR$-symmetric 
and if the SM quark doublet $Q=(t_L,b_L)$ couples with a composite quark in a $(2,2)_{2/3}$ of $\SU(2)_L\otimes \SU(2)_R \otimes \U(1)_X$ where $Y=T_R^3 + X$.
The embedding of $b_L$ into such a representation satisfies the condition above.

In the fundamental theory, 
this protection mechanism occurs automatically in 
the $\SO(N)_{\rm TC}$ model described above,
with TC-fermions $\Q= \Q_L\oplus  \Q_{L^c}\oplus  \Q_N$. In fact, if
 the mass difference between
$\Q_L$ and  $ \Q_{L^c}$ is much smaller than $\Lambda_{\rm TC}$
they form a left-right-symmetric bidoublet.
Adding a TC-scalar $\S_{U^c}$ allows to couple the SM quarks $Q$ and $U$ to the TC-particles, obtaining the top Yukawa couplings.
Then, the fermionic bound states ${\cal B}=\F\S$ contain 
an accidentally degenerate pair of composite fermions 
$\S_{U^c} \Q_{L} = (3,2)_{1/6}$
and
$\S_{U^c} \Q_{L^c} = (3,2)_{7/6}$,
where we showed the SM gauge quantum numbers.
By mixing with them, $Q$ keeps its SM value of the $Zb\bar b$ coupling, up to higher order corrections.
A model will be discussed in section~\ref{SOmodel}.
In the $\SU(N)_{\rm TC}$ and $\Sp(N)_{\rm TC}$ cases such a mechanism would require a more involved construction.

\section{Successful models}\label{models2}

Having discussed separately the main ingredients, we now study if concrete models exist that realize
simultaneously all the 4 following conditions:
\begin{itemize}
\item[1.] The new strong gauge interaction is asymptotically free and generates condensates.
For a given content of TC-particles, this implies an approximated lower bound on $N$, see eq.\eq{N>}. 

\item[2.] All couplings can be extrapolated up to the Planck scale without hitting 
Landau poles.  For the SM gauge couplings this implies that their one-loop $\beta$ function coefficients $b_i$ 
defined by 
$\sfrac{d \alpha_i^{-1}}{d \log E} = - \sfrac{b_i}{2 \pi} + \cdots$
must satisfy the conditions
\be\label{eq:blim}
b_3 \lesssim 1.9 \,, \qquad b_2 \lesssim 5.3 \,, \qquad b_1 \lesssim 10 
\ee
having assumed  $\Lambda_{\rm TC}\sim \TeV$ and written $b_1$ in $\SU(5)_{\rm GUT}$ normalization
(equivalently $b_Y = \frac53 b_1 \lesssim 16.6$).
For a given TC-particle content, these conditions imply an upper bound on $N$.

\listpart{The above two requirements are compatible if the TC-particle content is small enough.
However, the third condition requires a large enough TC-particle content.}

\item[3.]  Each generation of SM fermion $L,D,U,Q,E$ must acquire mass.
In effective scenarios, one requires that each SM fermion mixes with a composite state;
in our models this translates into Yukawa couplings involving a SM fermion, a TC-scalar and a TC-fermion. 
But this is not enough: it can still happen that some masses are either forbidden for symmetry reasons and/or that operators that break baryon and lepton number are generated.\footnote{An example of an unfortunate model plagued  by both problems is best described in $\SU(5)_{\rm GUT}$ language: the SM fermions are $\bar 5\oplus 10$,
the TC-particles are $\F_{\bar 5}\oplus \bar \F_1$ and $\S_5$,
the Yukawas are $\bar 5 \F_1^c \S_5+10\,\F_{\bar 5} \S_5^*$.
No up quark mass is generated.
Baryon number is violated because, like in any model that employs full SU(5) representations,
$Q$ and $U$ are both contained in the 10, but have opposite baryon number.}

\end{itemize}
The above conditions eliminate the most naive models\footnote{For example those with a TC-fermion for each SM fermion
and TC-scalars with the same SM gauge quantum numbers as the SM Higgs doublet, or
of a neutral singlet.} and require us to devise 
an economical enough set of TC-particles with the same accidental $\U(1)_{B,L}$ symmetries of the SM.
Quarks with equal baryon number and
leptons with equal lepton number can be combined in right-handed doublets 
$Q_R=\{ U,D\}$ and $L_R=\{E,N\}$, where $N$ is an optional right-handed neutrino.
The Yukawa couplings must have the generic form
\beq \Lag_Y \sim (Q \F \S^*_{q}+Q_R\F ^c \S_{q}) + ( L \F \S_\ell^* + L_R \F^c \S_\ell ) \eeq
such that the (so far unspecified) TC-particles $\F,\S_{q,\ell}$
mediate all SM Yukawa couplings.
This can be more easily
seen in the artificial limit where the TC-scalars $\S_{\ell,q}$ are so heavy that they can be integrated out at tree level
as in fig.~\ref{flavour},
giving rise to 4-fermions operators $LL_R\F\F^c + QQ_R \F\F^c$.
A similar structure arises if TC-scalars are below the confinement scale.
The TC-fermion bilinears necessarily have the quantum numbers of a Higgs doublet,
and do not contain any lepto-quark.

\smallskip
\indent

The fact that an $\SU(2)_R$ structure automatically emerges is beneficial for the last condition.

\begin{itemize}
\item[4.] The model must be compatible with experimental bounds.  
LHC bounds force the new particles to be heavier than about $\LTC\circa{>}1\TeV$.
Precision data (mostly the $T$ parameter and $Zb_L\bar b_L$) imply the stronger bound 
$\LTC\circa{>}5\TeV$, corresponding to a large fine-tuning ${\rm FT} \circa{>} 100$ in the Higgs mass.
One can follow different strategies, and it is not clear which one is preferable:
\begin{itemize}
\item[4a.] Accept a large fine-tuning.

\item[4b.] Build ad-hoc models aiming at suppressing the unnaturally large quantum corrections to the Higgs mass.

\item[4c.] Conceive models able to suppress corrections to $T$  and $Zb_L\bar b_L$, such that
$\LTC\circa{>}1\TeV$ corresponding to ${\rm FT} \circa{>} 10$, becomes allowed.
This can be realized through custodial symmetries that typically need a special TC-particle content.
\end{itemize}

\end{itemize}

\begin{figure}
$$\includegraphics{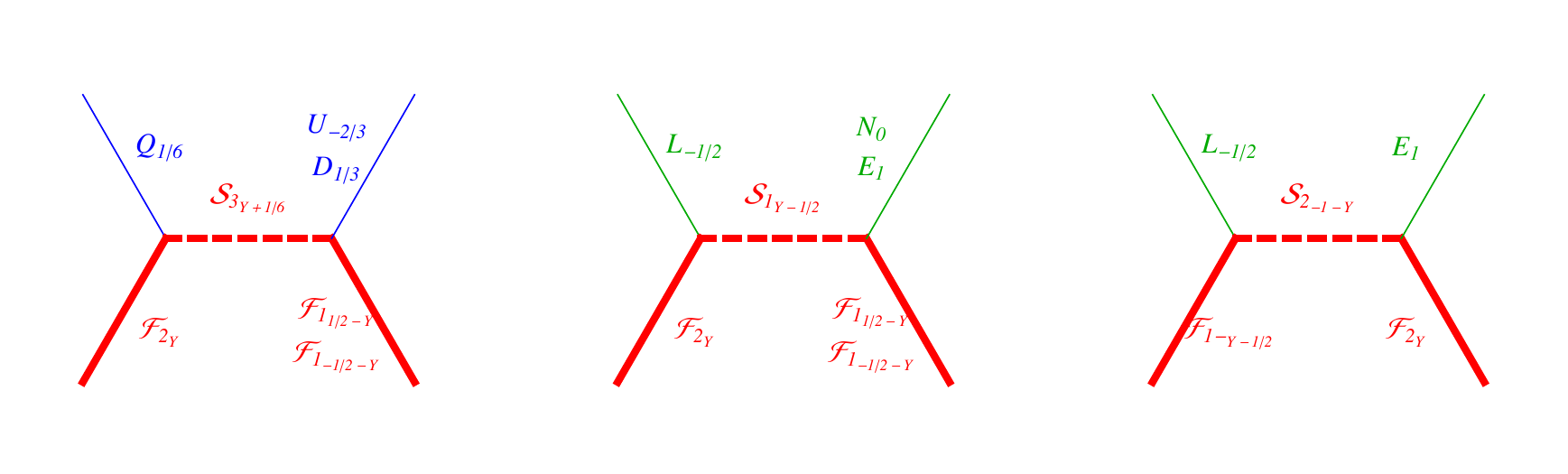}$$
\vspace{-1.8cm}
\caption{\em \label{AllowedModels}  Most general `economical' choice of quantum numbers.
The pedices in every field denote the hypercharge and $Y$ is a free constant.
3 denotes a color triplet, 2 a weak doublet and 1 a singlet.
Thick lines denote TC-particles; the presence of techni-strong interaction among them is understood.
}
\end{figure}

Figure~\ref{AllowedModels} shows the
most general `economical' choice of quantum numbers:
in the left diagram we assume that $Q$ couples to a TC-fermion weak doublet $\F_2$ 
(with generic hypercharge $Y$)
and to a TC-scalar color triplet $\S_3$.
Then $U$ and $D$ must couple to $\S_3$ and to TC-fermion singlets $\F_1$.
In the middle diagram, we next assume that $L$ couples to the same $\F_2$ and to a new TC-scalar singlet $\S_1$;
then $E$ and $N$ have hypercharges such that they
can couple to the $\F_1$ singlets previously introduced.
If $Y=\pm 1/2$, all TC-particles lie in $\SU(5)_{\rm GUT}$ fragments,
giving rise to the models described in sections~\ref{5frags2} and~\ref{5frags1}.
The minimal choice $Y=0$ gives the model in section~\ref{minimal}.

Alternatively, a less minimal model is obtained 
mediating lepton masses as described in the right diagram
(rather than as in the middle diagram): $E$ is coupled to $\F_2$ through a scalar doublet $\S_2$;
then $L$ can be coupled to one of the TC-fermion singlets $\F_1$.  Right-handed neutrinos $N$ remain uncoupled.
The model is described in section~\ref{Vigiani} for $Y=-1/2$.


\subsection{Model with $\SU(5)_{\rm GUT}$ fragments and $Y=-1/2$}\label{5frags2}
Referring to the first two diagrams of fig.~\ref{AllowedModels}, the choice $Y=-1/2$ corresponds to the TC-particle content
\beq N_{g_F} \times   (\F_{L} \oplus \F_{E^c}\oplus \F_N)\oplus N_{g_S} \times(
\S_{E^c}\oplus \S_{D^c}).\eeq
For extra clarity, table~\ref{tab:5frags2} lists the gauge quantum numbers of the TC-particles.
For $G_{\rm TC}=\SU(N)_{\rm TC}$ the most generic Yukawa couplings are 
\beq  \label{eq:Yuk2}
\Lag_Y =
y_L ~ L   \F_{L}  \S_{E^c}^*+
y_E ~ E \F_{N}^c \S_{E^c} +
(y_D ~ D \F_{N}^c +
y_U ~ U \F_{E^c}^c) \S_{D^c}+
y_Q ~ Q\F_{L}\S_{D^c}^*+
\hbox{h.c.} 
\eeq
The scalar interactions are
\beq
\lambda_{E} |S_{E^c}|^4+
\lambda_{E D}|S_{E^c}|^2\Tr(\S_{D^c}\S_{D^c}^\dagger)+
\lambda_{D} \Tr(\S_{D^c}\S_{D^c}^\dagger)^2+\lambda'_{D} \Tr(\S_{D^c} \S_{D^c} ^\dagger \S_{D^c} \S_{D^c} ^\dagger).\eeq
The renormalizable interactions conserve 5 accidental U(1) global symmetries.
First, an anomaly-free TC-baryon number, with charges equal to $+1$ ($-1$) for
TC-particles in the $N$ ($\bar N$), and to 0 for SM particles.
Next, baryon and lepton-number, under which the 
TC-fermions are neutral and the TC-scalars compensate for the charge of the SM fermions. 
Finally, two extra U(1) restrict the interactions among SM particles and TC particles.
No extra Yukawa couplings nor cubic scalar couplings are allowed for $N=3$.

Analogous Yukawa couplings can be written in the $\SO(N)_{\rm TC}$ ($\Sp(N)_{\rm TC}$) case, with the only difference that the fundamental is a real (pseudo-real) representation, {such that one can identify $\F_N=\F_N^c$ breaking TC-baryon number.
For $\Sp(N)_{\rm TC}$ two copies of the $\F_N$ singlet must be considered in order to avoid topological anomalies.}
The one-loop coefficient of the TC gauge $\beta$ function is 
\beq
b_{\rm TC} =\left\{\begin{array}{ll}
  -\frac{11}{3} N + \frac23 (4 N_{g_F} + N_{g_S} ) & {\rm for} \, \SU(N)_{\rm TC} \,, \\
 -\frac{11}{3} (N-2) +  \frac{2}{3} (7 N_{g_F} + 2 N_{g_S} )& {\rm for} \, \SO(N)_{\rm TC} \,,  \\
 -\frac{11}{6} (N+2) + \frac23 (4 N_{g_F} + N_{g_S} )   & {\rm for} \, \Sp(N)_{\rm TC} \,.  
 \end{array}\right.
\eeq
We verified that the two-loop term is subdominant for $g_{\rm TC}\circa{<} 4\pi$.
In all cases the one loop coefficients of the SM gauge $\beta$ functions are
\beq
b_{3} = -7 + \frac{N}{6} N_{g_S} ,\qquad
b_{2} = -\frac{10}{3} + \frac{2 N}{3} N_{g_F} \,, \qquad
b_{1} =  4 + \frac{6}{5} N N_{g_F} + \frac{4}{15} N N_{g_S}  .
\eeq
We assume $N_{g_F}=1$, $N_{g_S}=3$, which is the most economic choice that allow to write Yukawa couplings for all the 3 generations of 
SM fermions and to satisfy the conditions on the $\beta$ functions.
All gauge $\beta$ functions lie in the allowed range for $1.9\circa{<}N<3.0$ in the $\SU(N)_{\rm TC}$ case 
  and for
$1.8\circa{<}N<3.0$ in the $\Sp(N)_{\rm TC}$ case
(for $N=2$ one has $\SU(2)=\Sp(2)$).
No solution is found for $\SO(N)_{\rm TC}$.\footnote{No solutions are found for 
any group for the alternative choice $N_{g_F}=3$, $N_{g_S}=1$, which would also lead to a large set of TC$\pi$ made of two TC-fermions.
}

\begin{table}[t]
$$\begin{array}{|ccc|cccc|}\hline
\hbox{name} &\hbox{spin}  & \hbox{generations} & \SU(3)_c & \SU(2)_L & \U(1)_Y &G_{\rm TC}  \\
\hline 
\F_N &1/2& N_{g_F}& 1 & 1 & 0  &N  \\  
\F_N^c &1/2& N_{g_F}& 1 & 1 & 0  &\bar N  \\  
\F_{L} &1/2&  N_{g_F}&1& 2 & -1/2 &N   \\
\F_{L}^c &1/2&  N_{g_F}&1& 2 & +1/2 &\bar N   \\
\F_{E^c} & 1/2 &N_{g_F}& 1 &1 & -1 &  N\\
\F_{E^c}^c & 1/2 &N_{g_F}&  1 &1 & +1 & \bar N\\ \hline
\S_{E^c} &0& N_{g_S}& 1& 1 & -1 & N   \\
\S_{D^c}& 0 &N_{g_S} &  3 &1 & -1/3 &  N\\
\hline
\end{array}$$
\caption{\label{tab:5frags2}\em Explicit field content of the model of section~\ref{5frags2} in the case of $G_{\rm TC}=\SU(N)_{\rm TC}$.
}
\label{tab:most-minimal-model}
\end{table}


Table~\ref{tab:G2H} gives the global symmetry breaking pattern in the fermionic sector:
$\SU(8)\to \Sp(8)$ for  $\Sp(N)_{\rm TC}$
and $\SU(4)_L \otimes \SU(4)_R \to \SU(4)_V$ for  $\SU(N)_{\rm TC}$.
The latter possibility corresponds to the minimal one that {can} realize a custodial symmetry as described in point 2 of section~\ref{custodial}:
the chiral symmetry breaking produces 15 light TC$\pi$ in the adjoint of $\SU(4)_V$, that decomposes as
\beq
\hbox{TC}\pi =  2 \times (1,1)_0 \oplus (1,3)_0 \oplus [(1,1)_1 \oplus 2 \times (1,2)_{-1/2} + \hbox{h.c.} ]  \qquad\hbox{under $G_{\rm SM}$.}
\eeq
All TC$\pi$ are unstable.
Among these TC$\pi$ we can identify two Higgs doublets that can be embedded in a complex bidoublet of the unbroken custodial symmetry $\SU(2)_L \otimes \SU(2)_R$. 
As said above, in the presence of two Higgs doublets, generic vacuum expectation values break the custodial symmetry: 
in appendix~\ref{concreto} we show how a custodial preserving minimum can be obtained in this model (and more in general in $\SU(N)_{\rm TC}$ models).

TC-baryon number, conserved in $\SU(3)_{\rm TC}$ models, implies that the lightest TC-baryon is stable.
If TC-particle masses are such that the lightest TC-baryon is the neutral spin-$3/2$ TC-baryon $\F_N^3$,
it can be identified with Dark Matter, either as a thermal relic or with a TC-baryon asymmetry~\cite{1503.08749}.

\medskip

{Neutrino masses can be obtained  adding right-handed neutrinos $N$, which
can have  the Yukawa couplings
\beq y_N\,  N \F_{E^c}^c \S_{E^c} + y'_N N \,\F_{E^c} \S_{E^c}^*+\hbox{h.c.}\eeq
as well as Majorana masses $M_N$.
The Yukawa couplings $y_N$ together with $y'_N$ break lepton number and contribute to $M_N$ as $ \sim y_N y'_N \Lambda_{\rm TC}/g_{\rm TC}^2$.
If $M_N=0$ and $y'_N=0$ lepton number is conserved, and
neutrinos acquire Dirac masses $m_\nu \sim y_L y_N v/g_{\rm TC}$.
If instead $N$ have large Majorana masses $M_N$, they can be integrated out obtaining the dimension-5
effective operators $y_N^2(\F^c_{E^c} \S_{E^c})^2/M_N$ among TC-colored particles, {which breaks lepton number by 2 units and respects TC-baryon number}.
Integrating TC-particles out at the $\Lambda_{\rm TC}$ scale, taking into account the $y_L$ couplings, gives rise to Majorana
neutrino masses $m_\nu \sim (y_N y_L/g_{\rm TC})^2 v^2/M_N$.}

\subsection{Model with $\SU(5)_{\rm GUT}$ fragments and $Y=+1/2$}\label{5frags1}
For brevity, we only describe the main differences with respect to the previous model.
Setting $Y=+1/2$ we obtain the TC-particle content
\beq 
(\F_{L^c} \oplus \F_{E}\oplus \F_N)\oplus 3 \times(
\S_N\oplus \S_{U^c}). \eeq
The model has a built-in custodial symmetry, with the same TC$\pi$ content as the previous model.
For $G_{\rm TC}=\SU(N)_{\rm TC}$ the most generic Yukawa couplings are 
\beq  
\Lag_Y =
y_L ~ L   \F_{L^c}  \S_{N}^*+
y_E ~ E \F_{E}^c \S_{N} +
(y_D ~ D \F_{E}^c +
y_U ~ U \F_{N}^c) \S_{U^c}+
y_Q ~ Q\F_{L^c}\S_{U^c}^*+
\hbox{h.c.} 
\eeq
Gauge $\beta$ functions lie in the allowed range for $1.9\circa{<}N<3.0$ ($\SU(N)_{\rm TC}$ and for
$1.8\circa{<}N<3.0$ ($\Sp(N)_{\rm TC}$).
For $N=3$ the extra terms
$\F_N\F_N\S_N$ and (for $N_{g_S}\ge 3$ )  $S_N^3$ are allowed, 
breaking TC-baryon and lepton numbers.
As a consequence the lightest TC-baryon becomes unstable, and 
$\Delta L=3$ 4-fermion  interactions between 3 SM leptons and heavy composite fermions are generated.
These are not subject to the significant bounds that hold on $\Delta L=2$ effects.
If instead the extra terms are absent, the lightest TC-baryon of $\SU(3)_{\rm TC}$ is stable, and
it is a good Dark Matter candidate if made of $\S_N$ and/or $\F_N$. {Neutrino masses can be obtained,
similarly to section~\ref{5frags2}, adding right-handed neutrinos $N$ with Yukawa couplings $N \F_N^c \S_N$.
}

\subsection{Model with minimal custodial symmetry and $Y=0$}
\label{minimal}
The choice $Y=0$ does not correspond to TC-particles in fragments of $\SU(5)_{\rm GUT}$ but 
leads to an interesting model with simple representations
\beq
W\equiv  (1,2)_0,\qquad
Y \equiv (1,1)_{1/2},\qquad
X=(3,1)_{1/6}. \eeq
As discussed in section~\ref{custodial}, these are the representations that can realize the
minimal coset with a custodially protected Higgs doublet from $G_{\rm TC}=\Sp(N)_{\rm TC}$ dynamics.
The needed TC-particle content is:
\beq N_{g_F}\times(\F_{ W}\oplus \F_Y \oplus \F_{Y^c} )\oplus
N_{g_S}\times (
\S_{Y} \oplus \S_{ X})\,,\eeq
such that the allowed Yukawa couplings are
\beq \Lag_Y = 
y_L\,L \F_{W} \S_Y +
y_E\,E   \F_{Y^c}  \S_Y^*  +
(y_U\, U  \F_Y +
y_D\, D \F_{Y^c} ) \S_X + 
y_Q\, Q  \F_W \S_X^* .
\eeq
All composite states have non-exotic gauge quantum numbers.
Assuming $N_{g_F}=1$ and $N_{g_S}=3$ the $\beta$ functions are in the allowed range for 
$0.7\circa{<}N<14.9$\footnote{Solutions with $1.8\circa{<}N<8.6$ also exist for $N_{g_F}=3$ and $N_{g_S}=1$, but lead to multiple Higgs doublets.}:
this range is larger than in previous models thanks to fact that $\F_W$ is real.
{This model contains no stable TC-baryons.}
{Neutrino masses can be generated thanks to a  $N \F_Y \S_Y^*$ coupling.}



\smallskip

\subsection{Model with $\SU(5)_{\rm GUT}$ fragments and scalar doublet}\label{Vigiani}
Setting $Y=-1/2$ the less minimal model with the scalar doublet, outlined in the right-handed panel of fig.~\ref{AllowedModels},
corresponds to the TC-particle content
\beq N_{g_F}\times (\F_L\oplus \F_{E^c}\oplus \F_N)
\oplus N_{g_S}\times (\S_{L^c}\oplus\S_{D^c}).\eeq
It automatically contains a custodial symmetry.
For $G_{\rm TC}=\SU(N)_{\rm TC}$ the full set of Yukawa couplings is
\beq  
\Lag_Y =
y_L ~ L   \F_{N}^c  \S_{L^c}+
y_E ~ E \F_L \S_{L^c}^* +
(y_D ~ D \F_{N}^c +
y_U ~ U \F_{E^c}^c )\S_{D^c}+
y_Q ~ Q\F_L\S_{D^c}^*+
\hbox{h.c.} 
\eeq
Notice that integrating out $\S_{D^c}$ gives a $\sim Q(D+U)\F\F$ 4-fermion operator, which gives mass to both up and down-quarks;
integrating out $S_{L^c}$ gives lepton masses.   
The model contains lepto-quarks with masses of order of $\Lambda_{\rm TC}$
coupled to $\bar D \gamma_\mu L$ and to $\bar E \gamma_\mu Q$, while the previous models contained
lepto-quarks coupled to $\bar Q \gamma_\mu L$ and to $\bar D \gamma_\mu E$.

Models with $N_{g_S}=3$, $N_{g_F}=1$ have all
 $\beta$ functions  in the desired range 
for $G_{\rm TC}=\SU(N)_{\rm TC}$  with $2.1\circa{<}N< 3.5$, 
(for $N=3$ the extra Yukawa couplings $\F_N\F_L \S_{L^c} + \hbox{h.c.}$ are allowed
{making the lightest TC-baryon unstable}),
and in the unphysical range
$2.2\circa{<}N< 3.5$ for  $G_{\rm TC}=\Sp(N)_{\rm TC}$.
%
{As anticipated, in this model right-handed neutrinos remain uncoupled.}

\subsection{Imperfect $\SO(N)_{\rm TC}$ model with minimal custodial symmetries}\label{SOmodel}
$\SO(N)_{\rm TC}$ model realizing a minimal custodial symmetry (as described in section \ref{custodial}) can be obtained splitting the first diagram in fig.~\ref{AllowedModels} in two diagrams, one for $U$ and another for $D$, mediated by different TC-scalars.
The required Yukawa couplings involving the fundamental TC-states
\beq (\F_{L} \oplus \F_{L^c}\oplus \F_N)  \oplus 3\times(
\S_{U^c}\oplus \S_{D^c} \oplus \S_E)
\eeq
are
\beq  
\Lag_Y =
y_L ~ L   \F_L \S_E +
y_E ~ E \F_N \S_E^* +
y_D ~ D \F_N  \S_{D_c}+
y_U ~ U \F_N \S_{U_c}+
(y_Q ~ \F_{L^c} \S_{U_c}^* +
y'_Q ~ \F_L \S_{D_c}^* )Q+
\hbox{h.c.}
\eeq
This $\SO(N)_{\rm TC}$ model does not satisfy all the required conditions on the $\beta$ functions for $\SO(N)_{\rm TC}$.
Nevertheless,  it is worth discussing at least the less worse case, corresponding to $\SO(5)_{\rm TC}$: 
$\beta_{\rm TC}$ is negative, but the approximated condition of eq.\eq{N>} is not satisfied.\footnote{We insist with 3 generations of
TC-scalars, although 2 might be enough.}
Furthermore, the hypercharge gauge coupling hits a Landau pole around $10^{14} \GeV$.
Despite this problem, this model is interesting because it is the most economic $\SO(N)_{\rm TC}$ with a built-in Higgs bidoublet of the custodial symmetry. 
Moreover, the correction to the $Zb\bar b$ coupling is of order
$\delta g_{b_L} \sim y_Q^4 v^2/g_{\rm TC}^4 f_{\rm TC}^2$,
automatically protected from larger corrections 
of order $\delta g_{b_L} \sim y_Q^2 v^2/g_{\rm TC}^2 f_{\rm TC}^2$
thanks to a custodial symmetry along the lines of~\cite{continite},
as explained in section \ref{custodial}.
The pattern of global symmetry breaking is $\SU(5) \to \SO(5)$ corresponding to 14 TC$\pi$ including one Higgs bidoublet $(2,2)$ of $\SU(2)_L \otimes \SU(2)_R$ and triplets $(3,3)$ with vevs that can be set zero using mechanisms as in \cite{vecchi}.
{The lightest TC-baryon is stable thanks to a $\mathbb{Z}_2$ symmetry~\cite{1503.08749}, 
and could be a good Dark Matter candidate such as $\F_L \F_{L^c} \F_N^3$.}
{With this matter content, right-handed neutrinos are decoupled.}


\subsection{Model with a full family of TC-scalars}\label{TCpiscalar}
Finally, we present a model where the light Higgs boson is a TC$\pi$ made of
two TC-scalars.
We choose a minimal content of TC-fermions 
(three generations of neutral SM singlets) and one full family of TC-scalars
\beq 3\F_N
\oplus \S_{L}\oplus\S_{E^c}
\oplus \S_{U^c}\oplus\S_{D^c}\oplus \S_{Q}.\eeq
The $\beta$ functions lie in acceptable ranges for
$G_{\rm TC}=\SU(N)_{\rm TC}$ with $1.8\circa{<}N<8.9$,
$\Sp(N)_{\rm TC}$ with $1.7\circa{<}N<8.9$,
$\SO(N)_{\rm TC}$ with $5.6\circa{<}N<8.9$.
{Undesired scalar cubics or quartics are allowed for $N=3$ or $4$.
For larger $N$ the lightest TC-baryon is stable at renormalizable level, and can be an acceptable Dark Matter candidate.}

Each SM particle $f$ has a Yukawa coupling of the form
$ f\F_N \S_f^*$ or $f\F_N^c \S_{f^c}$.
Tree-level $\F_N$  exchange mediates $ff \S\S$ effective operators,
which give Yukawa couplings for the SM fermions $f$ after
identifying the Higgs doublet as $H\in \S\S$.

$H$ can be a light pseudo-Goldstone boson if a $\langle \S\S\rangle$ condensate
appropriately breaks the accidental global TC-flavor symmetry among TC-scalars.
Depending on the pattern of global symmetry breaking we can have one or more TC$\pi$ with the quantum number of a Higgs doublet. 
One can realize the custodial symmetry along the lines discussed in section \ref{custodial}.
For example, for an $\SU(N)_{\rm TC}$ group, one can add a TC-scalar singlet $\S_N$, obtaining the sector
$\S_L \oplus \S_{E^c} \oplus \S_N$.
Alternatively, a custodial symmetry is already present in the colored sector
$\S_Q \oplus \S_{D^c} \oplus \S_{U^c}$: the global symmetry contains $\SU(3)_c\otimes \SU(4)$
and the latter factor can get spontaneously broken to
$ \SU(2)_L \otimes \SU(2)_R$,
 leading to two custodially protected Higgs doublets.
{Neutrino masses can be generated adding a  TCscalar $\S_N$.}


\section{Higgs properties}\label{SMYuk}
Composite Higgs estimates  use  effective field theories descriptions  
that combine assumed patterns of symmetry breaking with dimensional analysis.
Having a fundamental theory featuring simultaneously a composite Higgs and partial compositeness, we now proceed to extract as much informations as possible.\footnote{State-of-the-art lattice simulations \cite{lattice} are providing vital informations on the pattern of chiral symmetry breaking, spin-one spectrum, decay constants, TC-fermion mass dependence, and scattering lengths for the minimal fundamental composite Higgs scenario \cite{1402.0233} discussed in subsection \ref{minimal}, but without TC-scalars. It would be interesting to investigate the dynamics of this theory including TC-scalars, in particular  to estimate the spectrum of composite baryons constituted by a TC-fermion and a TC-scalar.}

Chiral Lagrangians are tailored for  (pseudo) Goldstone bosons, here the TC$\pi$ states are indicated with $\Pi$.\footnote{To describe heavier states one would need to follow the prescriptions of the CCWZ formalism \cite{CCWZ}, that allows to include resonances that are lighter than the cutoff consistently with the symmetry of the system. This is, however, not enough to ensure that quantum corrections can be properly taken into account. Depending on the added massive states one can, for example, make use of the properly implemented  large N counting scheme \cite{Sannino:2015yxa}.  }
 In general, by integrating out the heavy states, one obtains a set of effective operators for the light fields.
We consider models where the Higgs doublet is a TC$\pi$ made of two TC-fermions, $\F\F^c$ if $G_{\rm TC}=\SU(N)_{\rm TC}$
and $\F\F$ otherwise.\footnote{We will leave this distinction implicit. Furthermore, we do not discuss
global symmetry breaking in the TC-scalar sector, which can lead
to extra TC$\pi$ made of two TC-scalars and described as $\S^{\ast}\S=f_{\rm TC}^2 {\cal U}_S$.}
We then have
\beq\label{eq:FF} \F\F = f_{\rm TC}^2 \Lambda_{\rm TC} \, {\cal U} ,\qquad
{\cal U} = \exp \frac{2i \Pi}{f_{\rm TC}}
\eeq
where  $\Lambda_{\rm TC}$ is the mass of unprotected composite states,
$f_{\rm TC} \sim \Lambda_{\rm TC}/g_{\rm TC}$ is the $\Pi$ decay constant, and
$g_{\rm TC}\sim 4\pi/\sqrt{N}$ is the estimated size of the coupling among composite states assuming a large-$N$ behaviour of the TC gauge theory.
{This applies to  $\F\F, \F\S, \S\S$ composite states, while TC-baryons have larger masses $\sim N \Lambda_{\rm TC}$. Notice that in models without TC-scalars partial compositeness needs a TC-baryon $\mathcal{B}$ and therefore the Higgs mass receives $\approx N$ times larger corrections than in our model where, instead, $\mathcal{B}=\F \S$ (see also section~\ref{sVH}).}

As outlined in fig.~\ref{AllowedModels}, the Yukawa couplings $f\F\S$ 
 induce $ff\F\F$ operators:
roughly speaking, $\F\F$ can be expanded as $\F\F\sim f_{\rm TC}^2 \Lambda_{\rm TC} + 2i f_{\rm TC} \Lambda_{\rm TC} {H} + \cdots$ where $H$ is the Higgs doublet,
leading to the $ffH$ SM Yukawa couplings, to be studied in section~\ref{sYuk}.
The Higgs potential is generated by $\F\F$ terms (present in our theory at tree level)
and by $\F\F\F\F$ terms (generated at higher orders), to be studied in section~\ref{sVH}.
Furthermore $ffff$ terms give rise to flavor effects, studied in section~\ref{flavor}.

\subsection{Yukawa couplings}\label{sYuk}
To estimate the Yukawa couplings we formally reduce the associated squared of the partial composite operator  $f {\cal B}\sim f\F\S$  to the more familiar 4-fermion operator $ff'\F\F$ as if it were mediated, at tree level, by the TC-scalars $\S$,
such that their coefficient is $y_{f} y_{f'}/m_\S^2$.\sfootnote{This approximation can be efficiently used also in the case of purely fermionic theories once all the relevant composite scalars have been identified group-theoretically.}
Of course, TC strong interactions are relevant and this estimate only captures the general properties of these operators. Nevertheless there is a limit in which this approximation  is precise (up to renormalization corrections) and corresponds to
TC-scalars heavier than the confinement scale, $m_\S\gg \Lambda_{\rm TC}$. However, at least for the top quark,  we need  $m_\S\circa{<} \Lambda_{\rm TC}$ and the TC-scalars take part in the strong dynamics.  Here the coefficient of  the $ff'\F\F$ operator is roughly estimated by setting the TC-scalar mass to be around $\Lambda_{\rm TC}$.  Finally, each resulting $ff'\F\F$ operator  can be 
 rewritten as $ff'\,  \Tr [\Pi_f \Pi_{f'} {\cal U}]$ with $\Pi_f$ the projector onto each SM fermion involved in the resulting Yukawa interactions. 
  
\medskip

For example the 4-fermion operator $\approx y_Q y_U QU \F\F/\Lambda_{\rm TC}^2$ generates
the SM top Yukawa coupling $y_t\, QUH$ with $y_t\approx y_U y_Q/g_{\rm TC}$.
In order to obtain $y_t \approx 1$, the underlying Yukawa couplings $y_U$, $y_Q$ must be large,
e.g.\ $y_U \sim g_{\rm TC}$, $y_Q\sim 1$.
We now show that such values are compatible with the fundamental TC dynamics, and indeed quite natural.

Let us consider a fundamental Yukawa operator $y_f \,f\S \F$, where $\S$ and $\F$ are TC-scalars and TC-fermions
in the fundamental $N$ of $G_{\rm TC}$ and
$f$ is a SM fermion with $n_f$ components.
The relevant RGE are\sfootnote{In a more formal language, such RGE describe the 
anomalous dimension of the composite operators  that mix with SM fermions
discussed in  partially composite works based on unknown dynamics.}
\beq (4\pi)^2 \frac{\partial g_{\rm TC}}{\partial\ln\mu}=bg_{\rm TC}^3,\qquad
 (4\pi)^2 \frac{\partial y_f}{\partial\ln\mu}=f_f y_f^3 -f_g g_{\rm TC}^2y_f,
 \eeq
 where
\beq 
f_f = \frac{{N+2n_f +1}}{2},\qquad f_g=6 C_N=6\left\{\begin{array}{ll}
(N^2-1)/2N & \hbox{for $G_{\rm TC}=\SU(N)$}\\
(N-1)/2 & \hbox{for $G_{\rm TC}=\SO(N)$}\\
(N+1)/4& \hbox{for $G_{\rm TC}=\Sp(N)$}\\
\end{array}\right. \  .
\eeq
The RGE flow has the IR-attractive pseudo-fixed point $y^2_f/g_{\rm TC}^2=(f_g+b)/f_f$,
such that $y_U$ ($n_f=3$) can be bigger than $y_Q$ ($n_f=6$).
Taking into account that $y_f\ll 1$ is the other pseudo-fixed point, Yukawa couplings can naturally be either very small
or of order $g_{\rm TC}$.

\bigskip

The top Yukawa gets enhanced if the composite fermion that mixes with the top is light,
and this possibility is often assumed in Composite Higgs scenarios based on effective descriptions.
However, in our fundamental theory all $\S\F$ composite fermions are expected to be quasi-degenerate
with a mass around $\LTC$,
 in view of the unbroken TC-flavor symmetries among fermions $\F$ and among scalars $\S$,
similarly to how the QCD nucleons have a mass around $\Lambda_{\rm QCD}$.\footnote{\label{foot:effettiva}
One can extend the chiral Lagrangian to approximatively include the interactions between light 
Goldstone bosons in ${\cal U}$ and some heavy composite states given that
all states interact respecting the global symmetries of table \ref{tab:symmetries},
in the limit where the small explicit breakings are neglected.
The interactions of the composite fermions ${\cal B}_{ai}=\mathcal{S}_a \mathcal{F}_i$ 
(where $a$ is a flavor index of the fundamental of $\U(N_S)$ and $i$ of the fundamental of $\SU(N_F)$)
are obtained as
\be
\mathcal{S}_a \mathcal{F}_i \sim f_{\rm TC}\, \mathcal{U}_{ij} {\cal B}_{aj}
\ee
in analogy to the QCD effective interactions of nucleons with pions.}
As in the QCD case, the degeneracy of TC-hadrons is broken by various effects:
\begin{itemize}
\item[$i)$] TC-particle masses give a correction of order
$\Delta M_{\cal B} \sim m_\F+m_\S$, where $m_\F$ and $m_\S$ are the possible constituent masses;

\item[$ii)$] SM gauge interactions give positive corrections of order
$\Delta M_{\cal B} \sim +\alpha_{\rm SM}\Lambda_{\rm TC}/4\pi$,
where $\alpha_{\rm SM}$ is the SM charge of each composite state.

\item[$iii)$]   Yukawa couplings can give larger corrections to the composite fermions that mix with the top quark.
This latter possibility might lead to lighter top-quark partners.

\end{itemize}

\subsection{Higgs potential}\label{sVH}\label{Sec:higgs}
The TC$\pi$ content is model dependent, and
the full set of TC$\pi$ can contain extra Higgs doublets and/or extra singlets.
We consider as SM-like Higgs the state that carries the vev $v$,
such that the physical Higgs boson $h$
contributes to the $W$ mass as  $M_W^2 =\frac{1}{2} g_2^2 f_{\rm TC}^2 \, \sin^2(h/{\sqrt{2}}f_{\rm TC})$, with $\langle h \rangle \sim 246\, \GeV$.
Since the Higgs is a pseudo-Goldstone boson, 
its potential is generated by interactions that break the accidental global fermionic symmetry
and can be parameterized by inserting symmetry-breaking terms in a symmetric expression written in terms of ${\cal U}$.
Like in the previous discussion, we need to consider three effects:
\begin{itemize}
\item[$i)$] The TC-fermion masses contribute as
\beq \label{Vm}
V_{m}=-c_m  f^2_{\rm TC} \Lambda_{\rm TC} \mathrm{Tr}[m_\F \mathcal{U}+\hbox{h.c.}].\eeq
where $m_\F$ is the TC-fermion mass matrix and $c_m$ is a ${\cal O}(1)$ coefficient, presumably positive like in QCD.
Specializing to the SM Higgs, it yields
\beq
V_{m}= -{2}c_m \, f_{\rm TC} ^2 \Lambda_{\rm TC} \sum_i m_{\Q_i} \cos\left(\frac{h}{{\sqrt{2}}f_{\rm TC}}\right)
\eeq
where the sum runs over the TC-fermions that make the Higgs. 
This term alone cannot break the electro-weak symmetry since it predicts $\cos(h/{\sqrt{2}}f_{\rm TC})=1$.
 \item[$ii)$] SM loops induce positive squared masses for the Higgs via the quantum-induced potential
\be \label{Vgauge}
V_{g}= - c_g \frac{3}{2(16\pi^2)} \Lambda_{\rm TC}^2 f_{\rm TC}^2 \left(g_2^2 \Tr[ \mathcal{U} T^a \mathcal{U}^\dag T^a] + g_Y^2 \Tr[ \mathcal{U} T_Y \mathcal{U}^\dag T_Y] \right)  + {\cal O}(g_{2,Y}^4)  
\ee
 where $T^a, T_Y$ are the generators of $\SU(2)_L\otimes\U(1)_Y$ and $c_g \approx \ln (4\pi/g_{\rm TC})>0$
is a positive ${\cal O}(1)$ coefficient. Specializing to the SM Higgs, it yields
\beq
V_{g}= c_g \frac{3(3g_2^2+g_Y^2)}{64\pi^2} \Lambda_{\rm TC}^2 f_{\rm TC}^2 \sin^2 \left(\frac{h}{{\sqrt{2}}f_{\rm TC}} \right) + {\cal O}(g_{2,Y}^4)  
\eeq
where we neglected subleading terms. Taken in isolation this term does not break the electro-weak symmetry.

\item[$iii)$] The Yukawa couplings give rise to effective $\F\F\F\F$ interactions stemming, for example, from  the first diagram of fig.~\ref{flavour}. The non trivial contribution comes when $Q$ and $U(D)$ are exchanged with an overall coefficient scaling like $\sim y_Q^2 y_{U(D)}^2$. The largest contribution comes from the top sector because of its large coupling $y_t$ yielding 
$V_y \sim N_c  (\bar\F^c y_U^\dagger y_U \F^c)(\bar \F y_Q^\dagger y_Q\F)/{(4\pi)^2m_\S^2}$.
Recalling eq.~\eqref{eq:FF}, this becomes
\beq  \label{potenziale-yt}
V_y 
  =- c_y  \frac{N_cy_Q^2 y_U^2f_{\rm TC}^4}{(4\pi)^2 }\Tr[\mathcal{U}{^\dag}\Pi_{Q}^\dag \Pi_{Q} \mathcal{U}  (\Pi_U^\dag \Pi_U)^* ]= - c_y  \frac{N_c y_Q^2 y_U^2f_{\rm TC}^4}{(4\pi)^2 }\sin^2 \left(\frac{h}{{\sqrt{2}}f_{\rm TC}} \right) .
\eeq
where $c_y$ is a (presumably) positive order one constant, and
$\Pi_{Q,U}$ are projectors over specific TC-fermions $Q,U$: their explicit expressions can be computed in each model,
see appendix~\ref{concreto}.
\footnote{\label{foot2}
Terms proportional to $y_{Q,U,D}^4$ do not arise if $G_{\rm TC}=\SU(N)_{\rm TC}$
because the TC$\pi$ are 
described by a ${\cal U}$ matrix that transforms as
$\mathcal{U}\to L \mathcal{U} R^\dagger$ under 
the $\SU(N_F)_L\otimes \SU(N_F)_R$ symmetry
(less formally, TC$\pi$ are $\F \F^c$ states).
The projectors $\Pi_{Q,U,D}$
(or, equivalently, the Yukawa couplings $y_{Q,U,D}$ written as matrices)
can be seen as spurions transforming as
$(\Pi_Q^\dagger \Pi_Q) \to L (\Pi_Q^\dagger \Pi_Q) L^\dagger$
and $(\Pi_{U,D}^\dagger \Pi_{U,D}) \to R^* (\Pi_{U,D}^\dagger \Pi_{U,D}) R^T$
such that eq.~\eqref{potenziale-yt} is the only invariant.
Terms proportional to $y_{Q,U,D}^4$ can arise if instead $G_{\rm TC}=\SO(N)$ or $\Sp(N)$,
because TC$\pi$ are $\F\F$ states.
More formally, they are described respectively by a symmetric  and anti-symmetric unitary matrix that transforms as 
$\mathcal{U}\to g \mathcal{U} g^T$ with $g\in \SU(N_F)$ 
such that contractions of the form 
$\Tr[\mathcal{U}(\Pi_{Q}^\dag \Pi_{Q})^T \mathcal{U}^\dag  (\Pi_Q^\dag \Pi_Q) ]$
are allowed.

Furthermore, some effective scenarios with symmetry structures not related to fundamental theories can have composite states in representations of the global group that lead to
 corrections to the potential quadratic (rather than quartic) in the $f{\cal B}$ mixing terms.
In our fundamental models $\F\F\F\F$ interactions quadratic in the Yukawa couplings are
generated by TC-penguin diagrams.
However such diagrams only lead to a constant contribution to the TC$\pi$ potential:
in our models the TC-fermions lie in the fundamental representation of $G_{\rm TC}$,
which implies that the only possible index contraction is $\propto {\cal U}^\dag {\cal U}={\mathbb I}$. The same results can be obtained with symmetry arguments as done above.}
This term alone cannot break the electro-weak symmetry.
\end{itemize}
Summing $V=V_m+V_g+V_y$ and expanding it as $-\frac12M_h^2 |H|^2 + \lambda_H|H|^4+\cdots$, we obtain
\bea
-M_h^2 &\sim&  c_m \left( \underset{i}{\sum} m_{\Q_i} \right)  \Lambda_{\rm TC} + \left( c_g \frac{3(3 g_2^2 + g_Y^2)}{64 \pi^2}  - c_y N_c \frac{ y_t^2}{16 \pi^2} \right)\Lambda_{\rm TC}^2 \, , \notag \\
\lambda_H &\sim&  \frac{c_y N_c y_Q^2 y_U^2}{12 (4\pi)^2} - \frac{c_g g_{\rm TC}^2 (3 g_2^2+g_Y^2)}{16 (4\pi)^2} \sim \frac{ y_t^2}{N} \,,
\eea
where in the second line we assumed $m_{\Q} \lesssim \Lambda_{\rm TC}$ 
and that the dominant contribution is given by $y_t$.
Notice that $V_g$ and $V_y$ acquire the form of SM loops, with SM couplings $g_2$, $g_Y$ and $y_t$,
with a naive cut-off at $\Lambda_{\rm TC}$. The electro-weak symmetry can be appropriately broken by a combination of the above effects, such that $M_h^2$ 
is positive and small.  
This tuning is possible since $V_m$ 
has a different functional dependence on $h$ with respect to $V_g$ and {especially} $V_y$.
These results are in line with  \cite{1402.0233}. 
In appendix~\ref{concreto} we  study in more detail the Higgs potential in the model of section~\ref{5frags2}.

\begin{table}\small
$$\begin{array}{|c|ccccc|cc|}\hline
\hbox{Coupling} & \multicolumn{5}{c|}{\hbox{Flavor symmetry of SM fermions}} &  \multicolumn{2}{c|}{\hbox{Flavor of TC-scalars}}\\
\hline  \hline       &\U(3)_L &\U(3)_E&\U(3)_Q&\U(3)_U&\U(3)_D&\U(3)_{\S_{E^c}}&\U(3)_{\S_{D^c}} \\ 
\hline  
\rowcolor[cmyk]{0.1,0,0.3,0}y_L & 3 &1&1&1&1&3&1  \\
\rowcolor[cmyk]{0.1,0,0.3,0}y_E & 1 &3&1&1&1&\bar 3&1  \\
\rowcolor[cmyk]{0.1,0,0.3,0}y_Q& 1 &1&3&1&1&1&3  \\
\rowcolor[cmyk]{0.1,0,0.3,0}y_U& 1 &1&1&3&1&1&\bar 3  \\
\rowcolor[cmyk]{0.1,0,0.3,0}y_D & 1 &1&1&1&3&1&\bar 3  \\
\hline 
\rowcolor[cmyk]{0,0,0.25,0}m^2_{\S_{E}}& 1 &1&1&1&1&3\otimes\bar 3&1  \\
\rowcolor[cmyk]{0,0,0.25,0}m^2_{\S_{D}} & 1 &1&1&1&1&1&3\otimes\bar 3   \\   \hline
\rowcolor[cmyk]{0,0.1,0.05,0}\lambda_{E}& 1 &1&1&1&1&(3\otimes\bar 3)^2&1  \\
\rowcolor[cmyk]{0,0.1,0.05,0}\lambda_{{D,D'}} & 1 &1&1&1&1&1&(3\otimes\bar 3)^2   \\ 
\rowcolor[cmyk]{0,0.1,0.05,0}\lambda_{ED} & 1 &1&1&1&1&3\otimes\bar 3&3\otimes\bar 3   \\ 
\hline
\end{array}$$
\caption{\em\label{tab:U3}  Transformation properties of the Yukawa couplings, scalar masses and scalar quartics
under flavor rotations of the 5 SM fermions and of the 2  TC-scalars.
For concreteness we considered those of the model of section~\ref{5frags2}, 
but  other models share the same flavor spurionic symmetric, for the reasons illustrated in fig.~\ref{AllowedModels}.
The $3\otimes\bar 3$ representation can be decomposed as adjoint plus singlet.
}
\end{table}

\subsection{Flavor violations}\label{flavor}
Explorations of flavor in Composite Higgs have been performed using effective theories~\cite{flavorPC}.
We here discuss flavor from the point of view of a fundamental theory.

Making flavor indices explicit,
we  can write the Yukawa matrices of each SM fermion $f=\{U,D,Q,L,E\}$
as $y_f^{ij}$ where $i$ runs over $N_g=3$ (number of  generations of SM fermions)
and $j$ runs over $N_{g_S}$ (number of generations of TC-scalars).
We assume the minimal choice $N_{g_S}=3$, $N_{g_F}=1$ and that there is
one Yukawa matrix per SM fermion $f$ (one can build models with more than one: see for example section~\ref{SOmodel} and the extended model of appendix~\ref{concreto}).

Then,  each $y_f$ can  be decomposed as $y_f = V_f^\dagger y_f^{\rm diag} U_f$,
where $V$ and $U$ are unitary matrices and $y_f^{\rm diag}$ is a diagonal matrix
with positive entries.
The 5 matrices $V_f$ can be rotated away, by redefining the SM fields as $f\to V_f f$. 
The 5 matrices $U_f$ are physical, provided that the 2 TC-scalars have non-trivial mass matrices $m_\S^2$.
If instead the {potential couplings among TC-scalars conserve flavor}, 
2 of the matrices $U_f$ can be rotated away,  leaving, for example, $U_D$, $U_U$ and $U_E$ as physical matrices.
The SM Yukawa couplings 
\beq-\Lag_Y^{\rm SM}=y_{LE} LEH^*+y_{QD} QD H^* + y_{QU} Q U H+\hbox{h.c.}\eeq
are obtained as described
in the middle panel of fig.~\ref{flavour}:
\beq \label{yukawa-flavor}y_{ff'} \approx 
y_f  \times t \Big(\frac{f_{\rm TC}\Lambda_{\rm TC}}{m_\S^2} \Big) \times y_{f'}^T = y_f^{\rm diag} U_f  \,  t\Big(\frac{f_{\rm TC}\Lambda_{\rm TC}}{m_{\S}^{2 \, \rm  diag}}\Big) \,
U_{f'}^T y_{f'}^{\rm diag}.\eeq
where the loop function equals $t(x)=x$  in the artificial limit $m_\S\gg\Lambda_{\rm TC}$.
If instead the explicit TC-scalars masses are below the compositeness scale we expect the major contribution to their mass to come from the underlying strong dynamics leading to the estimate $t(x)\approx 1/g_{\rm TC}$.
The CKM matrix results, as usual,  from the misalignment between $y_{QU}$ and $y_{QD}$.
The 2 or 4 extra flavor-violating matrices must have small enough mixing angles in order to satisfy 
flavor bounds. In particular these extra rotations act on the right-handed fermions, generating potentially dangerous operators not present in the SM.

In the SM, the Yukawa matrices $y_{QU}$, $y_{QD}$ and $y_{LE}$
can be conveniently seen as spurions under global $\U(3)^5$ rotations of
the 3 generations of $L,E,Q,U,D$ fields.
In the present models a common similar structure arises, as summarized in table~\ref{tab:U3}.
Specific models have specific patterns of TC-fermions $\F$, which further restrict the possible couplings
of the Higgs $H\sim \F\F$.

The spurionic structure significantly
restricts the form of the possible flavor effects~\cite{hep-ph/9610485}  and is similar enough to the
SM structure.

\subsubsection*{Electric dipoles and $\mu\to e\gamma$}
Electro-magnetic dipole operators contribute to electric dipole moments and to $\mu\to e\gamma$.
They arise at loop level from $ff\F\F V$ operators, which have the same spurionic structure
of the SM Yukawa couplings, and where $\F\F$ becomes the SM Higgs.

Dressing the tree level diagram that mediates lepton masses (middle of fig.~\ref{flavour}) with TC gluons and attaching a SM vector one obtains
\beq d_{LE}^{ij} (L_i \gamma_{\mu\nu} E_j) V_{\mu\nu},\qquad\hbox{with}\qquad
d_{LE}= \frac{ g_{\rm SM}  v }{g_{\rm TC}\Lambda_{\rm TC}^2}
y_L \cdot \tilde t \Big(\frac{\Lambda_{\rm TC}^2}{m_\S^2} \Big) \cdot y_E^T
\eeq
for leptons, with similar results for up and down quarks.
Here $v$ is the Higgs vev and
$\tilde t(x)$ is a loop function different from $t(x)$.
If its argument is a generic matrix, the dipole matrix $d_{LE}$ is not proportional to $y_{LE}$ 
and the electron dipole is $d_e\sim e m_e/\Lambda_{\rm TC}^2\sim 10^{-23}e\,{\rm cm}\, (\TeV/\Lambda_{\rm TC})^2$,
5 orders of magnitude above the bound $|d_e|<0.87~10^{-28}\ecm$~\cite{ACME}.
Similarly, if $d_{QU(D)}$ is not proportional to $y_{QU(D)}$ 
the electric and chromo-electric dipoles of light quarks $u,d$ 
give a neutron electric dipole $d_n \gtrsim e m_d/\Lambda_{\rm TC}^2\sim  10^{-22}e\,{\rm cm}  \,(\TeV/\Lambda_{\rm TC})^2$
much above the bound  $|d_n|< 2.9~10^{-26}e\,{\rm cm}$~\cite{dnexp}.

If instead $m_\S^2\propto {\mathbb I}$ (this can arise e.g.\ if TC-scalars have no mass term,
and acquire a mass of order $\Lambda_{\rm TC}$ from TC-strong interactions)
{and the TC-scalar potential conserves flavor},
the leading-order $d_{LE}$ becomes proportional to $y_{LE}$, such that 
it gives no flavor nor CP violation.
In such a case, effects only arise through higher order powers in the Yukawa couplings.
We assume that scalar quartics similarly conserve flavor.
A spurion analysis shows two possible effects.
One has the form
\beq d_{LE}\sim  \frac{  g_{\rm SM}}{g_{\rm TC}}\frac{v}{\Lambda_{\rm TC}^2}y_L \cdot X\cdot y_E^T
\qquad\hbox{with}\qquad X = \frac{( y_L^\dagger y_L)}{g_{\rm TC}^2 }, \frac{(y_E^\dagger y_E)^T}{g_{\rm TC}^2 }\eeq
which arises adding extra Yukawa loops on the TC-scalar propagator in the
middle diagram of fig.~\ref{flavour}.  
Assuming $y_L\sim y_E$ the estimated dipole $d_e$ gets reduced by $X_{ee}\sim y_e/g_{\rm TC}\sim 10^{-7}$,
becoming compatible with experimental bounds for $\Lambda_{\rm TC}\circa{>} 200\GeV$, for $g_{\rm TC}\sim 4\pi$. 
An analogous result 
applies to $d_n$ that becomes compatible with the limits for a similar scale $\Lambda_{\rm TC}$.
 
\medskip

A similar estimate can be done for  $\mu\to e \gamma$. The experimental bound
$\mathrm{BR}({\mu\to e\gamma})< 5.7\, 10^{-13}$~\cite{MEG} must  be compared
to the theoretical prediction
\beq \mathrm{BR}(\mu\to e\gamma)=(|d_{e\mu}|^2 + |d_{\mu e}|^2)\frac{384\pi^2 v^4}{m_{\mu}^2}\eeq
where, assuming $g_{\rm TC}\sim 4\pi$, $y_L\sim y_E$ and mixing angles $U_{e\mu}\sim \sqrt{m_e/m_\mu}$, one has 
$X_{\mu e, e \mu}\sim \sqrt{y_e y_\mu}/g_{\rm TC}$ and
$d_{e\mu,\mu e}\sim e X_{e\mu,\mu e}  \sqrt{m_e m_\mu}/\Lambda_{\rm TC}^2 $, 
from which we derive the bound $\Lambda_{\rm TC}\gtrsim 500\,\GeV$.
Similar considerations hold for $ff\F\F$ operators, which can give rise to flavor-violating Higgs decays.

The second higher order correction to the dipole matrix has the form
\beq d_{LE}\propto (y_L  y_E^T)T\eeq
where $T$ is a flavor trace arising from extra loops on the TC-fermion propagator.
An imaginary part in $T$ gives rise to EDMs, while $\mu\to e\gamma$ remains vanishing.
Using only the biggest Yukawa matrices one can have
\beq\label{ImT1}
\Im T = \frac{1}{g_{\rm TC}^{12}} \Im\Tr[(y_Q^\dagger y_Q)^2( y_U^\dagger y_U)^{T2}(y_Q^\dagger y_Q)( y_U^\dagger y_U)^T]\sim 
 \frac{y_t^4 y_c^2 V_{cb} V_{ub} V_{us} }{g_{\rm TC}^{6}} .
\eeq
Using  $y_D$ one can have
\beq\label{ImT2}
 \Im T\sim \frac{1}{g_{\rm TC}^6} \Im\Tr[(y_Q^\dagger y_Q)^T( y_U^\dagger y_U)( y_D^\dagger y_D)]\sim  \frac{y_t^2  y_b^2  V_{cb}^2}{g_{\rm TC}^2} .\eeq
Both contributions are safely small.

\subsubsection*{4-fermion operators}
New operators with 4 SM fermions have the Lorentz structure with the coefficient demanded by box diagrams like in the right panel of fig.~\ref{flavour}, and the flavor structure demanded also by spurionic considerations:
\beq \label{eq:4fbox}
\sim  \frac{(y_f^\dagger y_f)_{ij}(y_{f'}^\dagger y_{f'})_{i'j'}}{g_{\rm TC}^2 m_\S^2}
(\bar f_i \gamma_\mu f'_{j'})(\bar {f}'_{i'} \gamma_\mu f_j)\qquad\hbox{for any $f,f'=\{L,E,Q,U,D\}$}.\eeq
The extra operator 
$(\bar L \gamma_\mu Q)(\bar E \gamma_\mu D)$ appears in
the models of section~\ref{5frags2}, \ref{5frags1}, \ref{minimal}, 
while the extra operator 
$(\bar L \gamma_\mu D)(\bar E \gamma_\mu Q)$ appears
in the model of section~\ref{Vigiani}.
These operators can be thought as mediated by the lepto-quarks mentioned in section~\ref{Vigiani}~\cite{0910.1789}.
If one can ignore the fact that flavor contractions differ from those that define the SM Yukawa couplings
$y_{ff'}$, such coefficient is of order $y_{ff'}^2/\Lambda_{\rm TC}^2$,
having assumed $m_\S\sim\Lambda_{\rm TC}$.

Flavor data put the strongest  constraint on the 
$(\bar s_R d_L)(\bar s_L d_R)/\Lambda^2$ operator, which contributes to CP-violation in $K_0/\bar K_0$ mixing:
if complex it must be suppressed  by $|\Lambda|>3\times10^5\TeV$~\cite{boundsFlavor}.
In our scenario it arises as $\Lambda \circa{>}\Lambda_{\rm TC}/\sqrt{y_s y_d}\sim 10^4 \Lambda_{\rm TC}$.

TC-penguin diagrams contribute to $\Delta F=0,1$ processes by giving
extra operators of the form
$(\bar f \gamma_\mu y_f^\dagger y_f f)J$ where $J$ is a flavor-universal SM or TC current.
Bounds are weaker than those from $K_0/\bar K_0$. 
\section{Conclusions}\label{end}
We proposed renomalizable extensions of the Standard Model in which a new fundamental  gauge dynamics becomes strong at around a TeV and  yields a composite pseudo-Goldstone Higgs boson that gives mass to all SM quarks and leptons through partial compositeness.

  The first key ingredient in our construction is the simultaneous presence of fermions and scalars charged under the new strongly interacting gauge group, that allows to write Yukawa couplings to SM fermions (see for example eq.\eq{Yuk2}).
 The second is that the SM chiral fermions have specific gauge quantum numbers (i.e.\ specific hypercharges) that non-trivially allow to implement the partial compositeness scenario in these models.  This peculiarity is analogous to the SM case in which one Higgs doublet is enough to give mass to all fermions. Furthermore, right-handed SM fermions $(U,D)$ and $(E,N)$
have an $\SU(2)_R$-like structure that gets extended to new fermions with new strong interactions,
resulting in custodially-protected composite Higgses,  improving the agreement with data. 

Figure~\ref{flavour} shows how the SM-like Yukawa couplings to the composite Higgs arise at tree level, and how extra flavor violations arise at one loop level.
The renormalizable Yukawa couplings between one SM fermion, one TC-fermion and one TC-scalar feature a spurionic-like symmetry different from the structure
present in the SM, as summarized in table~\ref{tab:U3}.
Nevertheless, the structure is similar enough (e.g.\ flavor violations among quarks do not imply flavor violations among leptons), and
flavor bounds are satisfied if TC-scalars masses are flavor universal. This possibility can emerge if the bulk of the composite scalar mass comes from the fundamental dynamics, which is of the order of the compositeness scale. {However, since also the quartic couplings of TC-scalars can induce flavor violation, we have to assume they are flavor universal.}

In more detail, models based on  $\SU(N)_{\rm TC}$ are allowed for $N=2,3$; however the TC$\pi$ contain
two Higgs doublets such that their vacuum expectation values can break the custodial symmetry,
see sections~\ref{5frags2}, \ref{5frags1}, \ref{Vigiani} and appendix~\ref{concreto}.
Models based on $\SO(N)_{\rm TC}$ lead to a single custodially-protected Higgs; 
however they have Landau poles slightly below the Planck scale (section~\ref{SOmodel}).
Models based on $\Sp(N)_{\rm TC}$ (section~\ref{minimal})  lead to a single custodially-protected Higgs if one employs 
TC-particles in representations not compatible with $\SU(5)_{\rm GUT}$ unification.
Finally, models where TC$\pi$ are made of TC-scalars (section~\ref{TCpiscalar}) need assumptions about their strong dynamics.
{In some models the lightest TC-baryon is a stable Dark Matter candidate.}

\bigskip

In section~\ref{models} we discussed general features of fundamental composite theories with both gauged fermions and scalars, including their classical and quantum global symmetries, and possible patterns of dynamical symmetry breaking. We argued that the new class of composite theories we considered here offers viable and testable solutions to several currently unsolved problems plaguing composite extensions of the SM in search of a successful microscopic realization of  the partial compositeness scenario. Compositeness allows for the introduction of new TC-scalars and its non-perturbative dynamics can be further investigated via first principle lattice simulations.  {In this respect, we foresee some immediate advantages of `scalarphilic' fundamental theories because of the following concrete reasons: 
$a$) Adding TC-scalars on the lattice is less computationally demanding than adding TC-fermions; 
$b$) We do not rely on nearly-conformal dynamics, hence lattice computations do not need the usual extremely large volumes; 
$c$) The state-of-the-art simulations of SU(2) gauge theories \cite{lattice} can be readily extended with the TC-scalars needed in the model of section \ref{minimal}. This will jump-start the investigation of realistic composite theories from ab-initio computations. {Lattice simulations will also help to determine the spectrum and the phase diagrams of theories with TC-fermions and TC-scalars.\footnote{{The main difference with respect to theories featuring gauge-fermions for partial compositeness is that for these theories: a) the spectrum contains typically two different fermion representations; b) one still lacks an underlying realization able to link the composite baryons to the standard model fermions; c)   one generically requires  very large anomalous dimensions for the composite baryons that can only emerge in near-conformal field theories with highly non-trivial dynamics that, would they exist at all \cite{Pica:2016rmv}, are much harder to simulate on the lattice than the present theory. Last but not the least adding TC-scalars on the lattice it is an easier task than adding fermions.  One of the reasons is that chiral symmetry on the lattice for fermions is hard to maintain.}}   } 
} 
{Finally, from a `scalarphobic' standpoint, our theories can also be viewed},
at least at some intermediate energies,  as approximate descriptions of the yet to be found phenomenologically successful composite theories featuring only TC-fermions: here 
our scalars would be interpreted as intermediate composite states rather than elementary.

\footnotesize\small

\subsection*{Acknowledgements}
This work was supported by the grant 669668 -- NEO-NAT -- ERC-AdG-2014.  AT is supported by a Oehme Fellowship. AT thanks for hospitality the Aspen Center of Physics, which is supported by NSF grant PHY-1066293. The work of FS is partially supported by the Danish National Research Foundation grant DNRF:90. 
We thank Roberto Contino, Serguey Petcov, Alex Pomarol, Riccardo Rattazzi, Slava Rychkov and Michele Redi for useful discussions.
\appendix

\section{Basics of Sp($N$) groups}\label{Sp}
Sp($N$) is defined as the group of rotations $U= \exp(i T)=\exp(i \theta^a T_a)$ that leave
$\gamma=\varepsilon \otimes \mathbb I_{N/2}$ invariant, $U^T \gamma U = \gamma $, so that the generators satisfy
$ T_a^T \gamma + \gamma T_a = 0$.
In the canonical basis where the generators are hermitian, they can be written as block matrices:
\beq
T = \bigg( \begin{array}{cc} X + X^\dag & Y + Y^T \\ Y^\dag + Y^* & -(X^*+X^T)\end{array}\bigg)
\eeq
where $X+X^\dag$ is a $N/2 \times N/2$ complex hermitian matrix and $Y+Y^T$ is a complex $N/2 \times N/2$  symmetric matrix.
Therefore the dimension of the $\Sp(N)$ group is $\frac{1}{4}N^2 + \frac{1}{4} N(N+2)= \frac{N}{2}(N+1)$.
The $\Sp(N)$ generators $T_a$ in the fundamental are related to the complex conjugated $T_a^*$ as
\be
\label{pseudo}
-T_a^* = \gamma^{-1} T_a \gamma.
\ee
The representations $N$ and $N^*$ are not independent: $\tilde N\equiv \gamma N^*$ transforms as the fundamental $N$. 

\medskip

The kinetic term of $N_S$ complex scalars in the $N$ of $\Sp(N)_{\rm TC}$ has an enhanced accidental global symmetry $\Sp(2N_S)$.
To see this, let us start from the simplest case $N=2$ and $N_S=1$, namely one doublet of $\Sp(2) \sim \SU(2)$.
As well known, the Higgs kinetic term (neglecting the hypercharge) can be written in terms of a bi-doublet $\Phi = (H,\tilde H)$ with $\tilde H= \varepsilon H^*$ in an $\SU(2)_L \otimes \SU(2)_R$ invariant form. 
The global symmetry of the Higgs kinetic term is $\SU(2)_R \sim \Sp(2)$.
In the two Higgses case ($N=N_S=2$) one similarly finds a $\Sp(4)$ global symmetry~\cite{2HDM}. 
For general $N$ and $N_S$, one can construct a $N \times 2 N_S$ matrix $\Phi = ({\mathcal S}, \tilde {\mathcal S}) = ( \S _1, \tilde  \S _1, \dots , \S _{N_S}, \tilde  \S _{N_S})$ and write the scalar kinetic term as
\be
\Lag_{\rm kin} = \frac{1}{2} \Tr [ (D_\mu \Phi)^\dag ( D^\mu \Phi)] \,,
\ee
that is left invariant by a $2 N_S \times 2 N_S$ matrix acting as $\Phi \to \Phi M$.
The pseudoreality of $\Phi$
\be
\Phi^* = \gamma \Phi \gamma'   \,, \qquad  \gamma' = {\rm diag} (\varepsilon, \ldots , \varepsilon) \,,   
\ee
gives a condition on $M$:
\beq
\Phi^* \to \Phi^* M^* = \gamma \Phi \gamma' M^*  ,\qquad
\gamma \Phi \gamma' \to \gamma \Phi M \gamma' \,,
\eeq
which implies $\gamma' M^* = M \gamma'$, defining a $\Sp(2N_S)$ global symmetry analogously to eq.~\eqref{pseudo}.

\begin{table}
{\footnotesize
$$\begin{array}{c|c}
 \multicolumn{2}{c}{\hbox{Scalar vev }~\langle \S \rangle}\\
\hbox{Gauge group} & \hbox{Global group }  \\ \hline
\SU(N)_{\rm TC} \to\SU(N-1)_{\rm TC}& \U(N_S)\to\U(N_S-1) \\
\SO(N)_{\rm TC}\to\SO(N-1)_{\rm TC}& \mathrm{O}(N_S)\to\mathrm{O}(N_S-1) \\
\Sp(N)_{\rm TC}\to\Sp(N-2)_{\rm TC} &  \Sp(2N_S)\to\Sp(2N_S-2)
\end{array}$$
\caption{\label{tab:vev} \em  Minimal pattern of gauge and global symmetry breaking induced by one scalar vacuum expectation value.} }
\end{table}

\section{Higgs as a TC-scalar Goldstone boson}\label{altro}
In this appendix we further elaborate on the possibility (outlined in section~\ref{TCS?}) that 
the Higgs is an elementary pseudo-Goldstone boson, neutral under the unbroken part of $G_{\rm TC}$.

This can arise as follows.
The TC gauge couplings become larger at low energy, driving the quartic couplings
to negative values at an energy which can naturally be
not much above the confinement scale.
This triggers a vacuum expectation value for the TC-scalars through the Coleman-Weinberg mechanism.
Depending on which of the stability conditions discussed below eq.\eq{VS} is violated,
either one or $N$  TC-scalars acquire vacuum expectation values.

Let us assume that only one TC-scalar $\S _{11}$ acquires a
vacuum expectation value: it breaks  the gauge symmetry
(its unbroken part decouples from SM particles)
and the TC-flavor global symmetry as described in table~\ref{tab:vev},
leaving an approximated Goldstone boson 
in the fundamental of the broken TC-flavor group.
Yukawa couplings can explicitly break the TC-flavor symmetry, giving mass
to the pseudo-Goldstone boson.


We present two models where the elementary pseudo-Goldstone boson can be identified
with the Higgs boson.
In both cases the TC-particle content is so large that the $\beta$ functions never lie in the desired range, if we insist on reproducing the masses of all SM fermions:
we thereby focus only on third generation quarks, ignoring the other SM particles.

\medskip

The first model considers an $\SU(N)_{\rm TC}$ gauge theory with TC-particle content
\beq \F_{Q} \oplus 
\S_{L}\oplus \S_{L^c}\oplus \S_{N}.
\eeq
The $\beta$ coefficients lie in the desired range for $N=3$.
We assume that RGE corrections are dominated by TC effects, such that the dominant quartic couplings respect the
accidental global symmetry U(5).
As a consequence, the vev  $\langle \S_N\rangle=f_{\rm TC}$ (which   leaves $G_{\rm SM}$ intact) breaks $\SU(N)_{\rm TC}\to \SU(N-1)_{\rm TC}$ and 
$\U(5)\to \U(4)$, giving rise to approximate pseudo-Goldstone bosons, which fill
two Higgs bosons and one real pseudo-scalar. 
The global symmetry can be  explicitly broken by extra quartics not generated by $g_{\rm TC}$ and by Yukawas interactions 
\beq \Lag_Y = 
y_U\, U  \F_Q \S_{L}^*+
y_D\, D \F_{Q}  \S_{L^c}^* + 
y_Q\, Q  \F_Q^c \S_N +\hbox{h.c.} \,.
\eeq
The top Yukawa coupling arises as $y_t = y_U \sin\theta$,
where $\theta \sim f_{\rm TC} y_Q/M_\Q$ is the mixing between $\Q$ and the first component of $\F_Q$.

\medskip

The second model considers an  $\SO(N)_{\rm TC}$ gauge theory with TC-particle content 
\beq \F_{Q} \oplus \S_{L}\oplus  \S_{N}.
\eeq
The $\beta$ coefficients lie in the desired range for $N=6$.
This corresponds to the minimal coset SO(5)/SO(4) of~\cite{Agashe:2004rs}, such
that the pseudo-Goldstone boson is a single Higgs doublet.
The Yukawa couplings are
\beq \Lag_Y = 
y_U\, U  \F_Q \S_{L}^*+
y_D\, D \F_{Q}  \S_{L} + 
y_Q\, Q  \F_Q^c \S_N +\hbox{h.c.}  \,.
\eeq



\section{Detailed analysis of a model}
\label{concreto}

We here present explicit results for the model of section~\ref{5frags2}, although the same discussion applies to all models with $\SU(N)$ TC group. The TC-particle content is
\beq 
(\F_{L} \oplus \F_{E^c}\oplus \F_N)\oplus 3 \times(
\S_{E^c}\oplus \S_{D^c}), \eeq
for $G_{\rm TC}=\SU(N)_{\rm TC}$, so that the most generic Yukawa couplings are those of eq.~\eqref{eq:Yuk2}.
The conditions on the gauge $\beta$ functions are satisfied for $N=2,3$. 
We consider $N=3$ so that the pattern of global symmetry breaking is $\SU(4)_L \otimes \SU(4)_R  \to \SU(4)_V$, which has a $\SU(2)_L\otimes \SU(2)_R$ subgroup in the composite sector.\footnote{For $N=2$ the TC gauge group is $\SU(2) \sim \Sp(2)$ and the pattern of global symmetry breaking in the fermionic sector becomes $\SU(8)\to\Sp(8)$ giving rise to 27 TC$\pi$.}
The explicit 
embedding of $\SU(2)_L \otimes \U(1)_Y$ into $\SU(4)_V$ is:
\be
T^i_L = \frac12 \begin{pmatrix} 
\sigma^i & 0 \\
0 & 0 \end{pmatrix} \,, \qquad  T_Y =  \underbrace{\frac12 \begin{pmatrix} 
0 & 0 \\
0 & \sigma^3 \end{pmatrix} }_{T^3_R} + q_X,\, ~~~~\mathrm{where}~~~ q_X(\F_{L},\F_{N},\F_{E^c})=-\frac{1}{2}
\ee
and $\sigma_i$ are the Pauli matrices. Notice that $\U(1)_X$ is the non anomalous $\U(1)$ in the global symmetry $\U(4)_L \otimes \U(4)_R$.
The 15 pNGB in the adjoint of $\SU(4)_{V}$ can be decomposed under the SM as
\beq
15 = \underset{\phi_{\pm}}{\underbrace{((1,1)_1 + \hbox{h.c.})}} \oplus  \underset{\eta_1, \eta_2}{\underbrace{2 \times (1,1)_0}} \oplus \underset{\pi_\pm,\pi_0}{\underbrace{(1,3)_0}} \oplus ( \underset{H,H^\dag,H^{'},H^{' \dagger}}{\underbrace{2 \times (1,2)_{-1/2} + \hbox{h.c.}}} )   \,.
\eeq
The Goldstone matrix $\mathcal{U}=\exp(2i \Pi/f_{\rm TC})$
with $\Pi=\Pi^\dag$ has the explicit form
\be
\label{SU4matrix}
\Pi = \frac{1}{\sqrt{2}}
\begin{pmatrix} 
\frac12 (\eta_1 + \sqrt{2}\pi_0) & \pi_{+} &  H'_0 & 
H_+ \\
- & \frac12 (\eta_1 - \sqrt{2}\pi_0) &
H'_-  &  H_0 \\
- & - & \frac12 (- \eta_1 +\sqrt{2} \eta_2) & \phi_+ \\ 
- & - & -  & \frac12 (- \eta_1 -\sqrt{2} \eta_2) \end{pmatrix} .
\ee
From the kinetic term $(f^2_{\rm TC}/4)\Tr[(D_\mu {\cal U})^\dag (D^\mu {\cal U})]$ we get the $W$ and $Z$ masses,
finding that the vacuum expectation values of the two Higgs doublets $H \sim \F_{L} \F_{E^c}^c$ and $H' \sim \F_{L} \F_N^c$
contribute at tree level to the $T$ parameter as\footnote{Defining the complex bidoublet $M = (H', H)$,
our parametrization is equivalent to \cite{ferretti}, once $M$ is identified with $\Phi_1 + i\Phi_2$, where $\Phi_{1,2}$ are two real bi-doublets of $\SU(2)_L\otimes \SU(2)_R$. Moreover, we can rotate $H_0$ and $H_0'$ to the basis where only one doublet gets a vacuum expectation value, and our formul\ae{} take the form of section \ref{Sec:higgs} with the physical SM-like Higgs $h\equiv\mathrm{Re}\,(H_0+H_0')$.}
\be
\hat T = \frac{M_{W_3}^2}{M_{W_1}^2}-1= \frac{ \big[ \cos (\sfrac{\sqrt{2}|H_0|}{f_{\rm TC}} )-\cos (\sfrac{\sqrt{2}|H_0'|}{f_{\rm TC}}) \big]^2}{2 \big[\cos (\sfrac{\sqrt{2}|H_0|}{f_{\rm TC}} )\,\cos (\sfrac{\sqrt{2}|H_0'|}{f_{\rm TC}}) -1 \big]} ,
\ee
that vanishes if $|H_0|=|H_0'|$.
A sizeable misalignment between $H_0$ and $H'_0$ would give a contribution of order 
$v^2/f_{\rm TC}^2$  such that the experimental bound $|\hat{T}|\lesssim 2\times10^{-3}$ would imply $f_{\rm TC} \circa{>} 5 \TeV$.
A symmetry of the fundamental 
Lagrangian can protect the $T$ parameter from large corrections aligning $H_0$ and $H'_0$.  
All the interactions that generate the potential such as the Yukawa couplings must respect this symmetry or the vacuum will misalign $H_0\neq H'_0$.
The model that we are considering does not enjoy such a symmetry since the doublet $H \sim \F_{L} \F_{E^c}^c$ is only coupled to the top and $H' \sim \F_{L} \F_N^c$ only to the bottom.
On the other hand, TC-fermion masses and gauge interactions align the vacuum in a direction $|H_0|=|H_0'|$:
\begin{itemize}
\item[$i)$] The mass matrix for the TC-fermion masses in the custodial limit is 
\be
m_\F=\diag(m_L,m_L,m_R,m_R)
\ee
and from eq.~\eqref{Vm} we get the potential 
\be
V_m = - 2 c_m \Lambda_{\rm TC} f_{\rm TC}^2 (m_L + m_R)\bigg(\cos \frac{{\sqrt{2}}|H_0|}{f_{\rm TC}} + \cos \frac{{\sqrt{2}}|H'_0|}{f_{\rm TC}} \bigg)
\ee
that is symmetric under $|H_0| \leftrightarrow |H'_0|$.

\item[$ii)$] Gauge interactions contribute to the potential as in eq.~\eqref{Vgauge} giving
\bea
V_g &=& - \frac{ 3 c_g f_{\rm TC}^2 \Lambda_{\rm TC}^2 }{128 \pi^2} \bigg[ 
g_Y^2 \bigg( 6 - \sin^2 \frac{{\sqrt{2}}|H_0|}{f_{\rm TC}} - \sin^2 \frac{{\sqrt{2}}|H'_0|}{f_{\rm TC}}  \bigg)+ \notag \\
&&
+g_2^2 \bigg( 2 - \sin^2 \frac{{\sqrt{2}}|H_0|}{f_{\rm TC}} -  \sin^2 \frac{{\sqrt{2}}|H'_0|}{f_{\rm TC}}  + 4 \cos \frac{{\sqrt{2}}|H_0|}{f_{\rm TC}} \cos \frac{{\sqrt{2}}|H'_0|}{f_{\rm TC}} \bigg) \bigg]
\eea
that is again symmetric under $H_0 \leftrightarrow H'_0$.
Taking only the terms above, one can show that the vacuum is aligned in a direction that preserve the electro-weak symmetry.

\item[$iii)$] EW symmetry breaking is triggered by the contribution from the Yukawa couplings. The dominant contribution comes from the top (see eq.~\eqref{potenziale-yt}) and since the top couples only to $H$, it is not symmetric in $|H_0|,\, |H'_0|$
\be
V_y = - \frac{c_y N_c y_Q^2 y_U^2 f_{\rm TC}^4}{(4 \pi)^2} \, \sin^2  \frac{{\sqrt{2}}|H_0|}{f_{\rm TC}}  \,,
\ee
so that the minimum does not preserve custodial symmetry.
\end{itemize}
This can be cured by adding one TC-scalar $\S_U$ coupled to the third generation quarks, so that the extra Yukawa couplings are allowed
\be
\label{custodial-T}
\Delta \Lag_Y = (y'_D ~ D \F_{E^c} + y'_U ~ U \F_N) \S_U^* + y'_Q ~ Q \F_{L}^c \S_U \,.
\ee 
In the symmetric limit $y=y'$, the Lagrangian enjoys a discrete symmetry $\F \leftrightarrow \F^c$ and $\S_{D^c}^*\leftrightarrow \S_U$
(see \cite{2HDM} for the role of discrete symmetries in other models with two composite Higgs doublets). With this addition the $\beta$ functions lie in the desired range.
Focussing on the third generation quark sector, the Yukawa couplings of the SM fermions to the TC$\pi$ are
\beq
\Lag_Y^{\rm SM} = y_{QU} Q \Pi_Q \mathcal{U} \Pi_U^T U + y_{QD} Q \Pi_Q \mathcal{U} \Pi_D^T D  
+y'_{QU} Q \Pi'_Q \mathcal{U}^T \Pi_U^{'T} U + y'_{QD} Q \Pi'_Q \mathcal{U}^T \Pi_D^{'T} D 
+ \hbox{h.c.}
\eeq
with  $\Pi_{Q,U,D}$, $\Pi'_{Q,U,D}$ defined as
\bea
\label{eq:Pis}
&&\Pi_Q = \bigg(\begin{array}{cccc} 0 & 1 & 0 & 0 \\ -1 & 0 & 0 & 0\end{array} \bigg),\,\qquad 
\Pi_{U} = \big(\begin{array}{cccc}0 & 0 & 0& 1\end{array}  \big), \,\qquad 
\Pi_{D} =  \big(\begin{array}{cccc}0 & 0 & 1& 0\end{array}  \big) \,, \notag \\
&&\Pi'_Q = \bigg(\begin{array}{cccc} 1 & 0 & 0 & 0 \\ 0 & 1 & 0 & 0\end{array} \bigg),\,\qquad 
\Pi'_{U} = \big(\begin{array}{cccc}0 & 0 & -1 & 0\end{array}  \big), \,\qquad 
\Pi'_{D} =  \big(\begin{array}{cccc}0 & 0 & 0& 1\end{array}  \big) \,.
\eea
We can assign spurionic transformation properties under the global symmetry $\SU(4)_L \otimes \SU(4)_R$
\beq\begin{array}{ll}
\Pi_Q^\dag \Pi_Q \to L \Pi_Q^\dag \Pi_Q L^\dag \,, \quad  &\Pi_{U,D}^\dag \Pi_{U,D} \to R^* \Pi_{U,D}^\dag \Pi_{U,D} R^T  \,, \notag \\
{\Pi'}_Q^\dag \Pi'_Q \to R^* {\Pi'}_Q^\dag \Pi'_Q R^T \,, \quad& {\Pi'}_{U,D}^\dag \Pi'_{U,D} \to L {\Pi'}_{U,D}^\dag \Pi'_{U,D} L^\dag  \,, \notag \\
{\Pi'}_Q^\dag \Pi_Q \to R^* {\Pi'}_Q^\dag \Pi'_Q L^\dag \,, \quad& {\Pi'}_{U,D}^\dag \Pi_{U,D} \to L {\Pi'}_{U,D}^\dag \Pi'_{U,D} R^T  \,, 
\end{array}
\eeq
and, recalling that ${\cal U} \to L {\cal U} R^\dag$, we can construct all the possible invariants contributing to the potential.
Taking into account that $m_b \ll m_t$ the relevant terms are
\bea
 y_Q^2 y_U^2  \Tr [ (\Pi_Q^\dag \Pi_Q ){\cal U} (\Pi_U^\dag \Pi_U)^* {\cal U}^\dag ] &=& y_Q^2 y_U^2  \sin^2  \frac{{\sqrt{2}}|H_0|}{f_{\rm TC}} \,, \notag \\
y^{\prime 2}_Q y^{\prime 2}_U  \Tr [( {\Pi'}_Q^\dag {\Pi'}_Q) {\cal U}^T ({\Pi'}_U^\dag {\Pi'}_U)^T {\cal U}^* ] &=&y_Q'^2 y_U'^2  \sin^2  \frac{{\sqrt{2}}|H_0'|}{f_{\rm TC}} \,,  \\
y'_Qy_Q y'_U y_U  \Tr [( \Pi^{\prime\dagger}_Q {\Pi}_Q ) {\cal U} ({\Pi'}_U^\dag {\Pi}_U)^T {\cal U}^* ] + {\rm h.c.} &=& -2 y'_Qy_Q  y'_U y_U \cos(\phi_0+\phi_0') \sin  \frac{{\sqrt{2}}|H_0|}{f_{\rm TC}} \sin  \frac{{\sqrt{2}}|H_0'|}{f_{\rm TC}} \, , \notag 
\eea
where $H^{(')}_0=|H^{(')}_0|e^{i\phi^{(')}_0}$. These contributions arise from the `square' of the diagram contributing to the top mass. 
Only the latter term depends on the phases $\phi_0$, $\phi'_0$,
and the minimum corresponds to $\cos(\phi_0+\phi_0')=-1$.
The symmetry $y\leftrightarrow y'$ implies the alignement $|H_0|=|H_0'|$.  The symmetry allows also for other terms:
\bea
&& y_Q^2 y^{\prime 2}_Q  \Tr [( \Pi_Q^\dag \Pi_Q) {\cal U} ({\Pi'}_Q^\dag {\Pi'}_Q)^* {\cal U}^\dag ]= y_Q^2 y^{\prime 2}_Q \bigg[\cos^2  \frac{{\sqrt{2}} |H_0|}{f_{\rm TC}}+\cos^2 \frac{{\sqrt{2}} |H_0'|}{f_{\rm TC}} \bigg]  , \notag \\
&&y_Q^2 y^{\prime 2}_Q \Tr [ (\Pi_Q^\dag {\Pi'}_Q) {\cal U}^T ({\Pi'}_Q^\dag \Pi_Q)^T {\cal U}^\dag ] =-2 
y_Q^2 y^{\prime 2}_Q \cos  \frac{{\sqrt{2}} |H_0|}{f_{\rm TC}}\cos \frac{{\sqrt{2}}  |H_0'|}{f_{\rm TC}}  .
\eea
When added together  (as dictated by the symmetry of the strong dynamics), the sum does not
contribute to quadratic terms in the fields $H$ and $H'$.
Finally, we can write the invariants
\beq
 y_U^2 y^{\prime 2}_U  \Tr [( \Pi_U^\dag \Pi_U ){\cal U}^T ({\Pi'}_U^\dag {\Pi'}_U)^* {\cal U}^* ],\qquad 
 y_U^2y^{\prime 2}_U  \Tr [ (\Pi_U^\dag {\Pi'}_U) {\cal U} ({\Pi'}_U^\dag \Pi_U)^T {\cal U}^* ] ,
\eeq
that only contribute to the potential of the (broken) $\SU(2)_R$ triplet $\phi^\pm,\eta_2$.

\end{document}